\documentclass[12pt]{article}
\usepackage{graphicx}
\usepackage{lscape}
\usepackage{citesort}

\begin{document}

\def\ds{\displaystyle}
\def\beq{\begin{equation}}
\def\eeq{\end{equation}}
\def\bea{\begin{eqnarray}}
\def\eea{\end{eqnarray}}
\def\beeq{\begin{eqnarray}}
\def\eeeq{\end{eqnarray}}
\def\ve{\vert}
\def\vel{\left|}
\def\ver{\right|}
\def\nnb{\nonumber}
\def\ga{\left(}
\def\dr{\right)}
\def\aga{\left\{}
\def\adr{\right\}}
\def\lla{\left<}
\def\rra{\right>}
\def\rar{\rightarrow}
\def\nnb{\nonumber}
\def\la{\langle}
\def\ra{\rangle}
\def\ba{\begin{array}}
\def\ea{\end{array}}
\def\tr{\mbox{Tr}}
\def\ssp{{\Sigma^{*+}}}
\def\sso{{\Sigma^{*0}}}
\def\ssm{{\Sigma^{*-}}}
\def\xis0{{\Xi^{*0}}}
\def\xism{{\Xi^{*-}}}
\def\qs{\la \bar s s \ra}
\def\qu{\la \bar u u \ra}
\def\qd{\la \bar d d \ra}
\def\qq{\la \bar q q \ra}
\def\gGgG{\la g^2 G^2 \ra}
\def\q{\gamma_5 \not\!q}
\def\x{\gamma_5 \not\!x}
\def\g5{\gamma_5}
\def\sb{S_Q^{cf}}
\def\sd{S_d^{be}}
\def\su{S_u^{ad}}
\def\ss{S_s^{??}}
\def\sbp{{S}_Q^{'cf}}
\def\sdp{{S}_d^{'be}}
\def\sup{{S}_u^{'ad}}
\def\ssp{{S}_s^{'??}}
\def\sig{\sigma_{\mu \nu} \gamma_5 p^\mu q^\nu}
\def\fo{f_0(\frac{s_0}{M^2})}
\def\ffi{f_1(\frac{s_0}{M^2})}
\def\fii{f_2(\frac{s_0}{M^2})}
\def\O{{\cal O}}
\def\sl{{\Sigma^0 \Lambda}}
\def\es{\!\!\! &=& \!\!\!}
\def\ap{\!\!\! &\approx& \!\!\!}
\def\ar{&+& \!\!\!}
\def\ek{&-& \!\!\!}
\def\kek{\!\!\!&-& \!\!\!}
\def\cp{&\times& \!\!\!}
\def\se{\!\!\! &\simeq& \!\!\!}
\def\eqv{&\equiv& \!\!\!}
\def\kpm{&\pm& \!\!\!}
\def\kmp{&\mp& \!\!\!}


\def\simlt{\stackrel{<}{{}_\sim}}
\def\simgt{\stackrel{>}{{}_\sim}}


\title{
         {\Large
                 {\bf
P--odd asymmetries in polarized
$\Lambda_b \rar \Lambda \ell^+ \ell^-$ decay
                 }
         }
      }

\author{\vspace{1cm}\\
{\small T. M. Aliev \thanks
{e-mail: taliev@metu.edu.tr}~\footnote{permanent address:Institute
of Physics,Baku,Azerbaijan}\,\,,
M. Savc{\i} \thanks
{e-mail: savci@metu.edu.tr}} \\
{\small Physics Department, Middle East Technical University,
06531 Ankara, Turkey} }

\date{}

\begin{titlepage}
\maketitle
\thispagestyle{empty}

\begin{abstract}
We calculate various P--odd asymmetries appearing in the differential decay
width for the cascade decay $\Lambda_b \rar \Lambda(\rar a+b) \, V^\ast
(\rar \ell^+ \ell^-)$ with polarized and unpolarized heavy baryons including
new vector type interactions and using the helicity amplitudes. It is
obtained that the study of P--odd asymmetries can serve a good test for
establishing new physics beyond the SM.
\end{abstract}

~~~PACS numbers: 12.60.--i, 13.30.--a. 13.88.+e
\end{titlepage}

\section{Introduction}

Rare B--decays induced by the flavor--changing neutral current (FCNC) 
$b \rar s$ or $b \rar d$ transitions occur at loop level in the standard 
model (SM), since FCNC transitions that are forbidden in the SM at tree 
level provide consistency check of the SM at quantum level.
Moreover, these decays are also quite sensitive to the existence of new
physics beyond the SM, since new particles running at loops can give
contributions to these decays. New physics appear in rare
decays through the Wilson coefficients which can take values different from
their SM counterpart or through the new operator structures in an effective
Hamiltonian (see for example \cite{R7201} and references therein).

Among the hadronic, leptonic and semileptonic decays, the last decay channels 
are very significant, since they are theoretically, 
more or less, clean, and they have relatively larger branching ratio. 
With the help of the semileptonic decays $B \rar M \ell^+ \ell^-$ ($M$ being
pseudoscalar or vector mesons) described by the 
$b \rar s(d) \ell^+ \ell^-$ transition, one can study  
many observables like forward--backward asymmetry ${\cal A}_{FB}$,
lepton polarization asymmetries, etc. Existence of these observables is very
useful and serve as a testing ground for the standard model (SM) and in
looking for new physics beyond th SM. For this reason, many processes, like
$B \rar \pi(\rho) \ell^+ \ell^-$ \cite{R7202},
$B \rar K \ell^+ \ell^-$ \cite{R7203} and
$B \rar K^\ast \ell^+ \ell^-$
\cite{R7204,R7205,R7206,R7207,R7208,R7209,R7210,R7211} have been studied 
comprehensively. 

Recently, BELLE and BaBar Collaborations announced the
following results for the branching ratios of the
$B \rar K^\ast \ell^+ \ell^-$ and $B \rar K \ell^+ \ell^-$ decays:
\bea
{\cal B}(B \rar K^\ast \ell^+ \ell^-) = \left\{ \begin{array}{lc}
\left( 11.5^{+2.6}_{-2.4} \pm 0.8 \pm 0.2\right) \times
10^{-7}& \cite{R7212}~,\\ \\
\left( 0.78^{+0.19}_{-0.17} \pm 0.12\right) \times
10^{-6}& \cite{R7213}~,\end{array} \right. \nnb
\eea
\bea
{\cal B}(B \rar K \ell^+ \ell^-) = \left\{ \begin{array}{lc}
\left( 4.8^{+1.0}_{-0.9} \pm 0.3 \pm 0.1\right) \times
10^{-7}& \cite{R7212}~,\\ \\
\left( 0.34  \pm 0.07 \pm 0.12\right) \times
10^{-6}& \cite{R7213}~.\end{array} \right. \nnb
\eea
Another exclusive decay which is described at inclusive level by the 
$b \rar s \ell^+ \ell^-$ transition is the baryonic 
$\Lambda_b \rar \Lambda \ell^+ \ell^-$ decay. Unlike mesonic decays, the
baryonic decays 
could maintain the helicity structure of the effective Hamiltonian for 
the $b \rar s$ transition \cite{R7214}. Radiative and semileptonic decays of
$\Lambda_b$ such as $\Lambda_b \rar \Lambda \gamma$, $\Lambda_b \rar
\Lambda_c \ell \bar{\nu}_\ell$, $\Lambda_b \rar \Lambda \ell^+ \ell^-$
$(\ell = e, \mu, \tau)$ and $\Lambda_b \rar \Lambda \nu \bar{\nu}$ have been
extensively studied in the literature 
\cite{R7215,R7216,R7217,R7218,R7219,R7220}. More
details about heavy baryons, including the experimental prospects, can be
found in \cite{R7221,R7222}.

Many experimentally measurable quantities such as branching ratio
\cite{R7223}, $\Lambda$ polarization and single-- and double--lepton 
polarizations, as well as forward--backward asymmetries,
have already been studied in \cite{R7224,R7225} and 
\cite{R7226}, respectively. Analysis of such quantities can be useful for 
more precise determination of the SM parameters and in looking for new physics 
beyond the SM. 

In the present work we analyze the possibility of searching for new physics
in the baryonic $\Lambda_b \rar \Lambda \ell^+ \ell^-$ decay by studying
different P--odd asymmetries that characterize the angular dependence of
the angular decay distributions, with the inclusion of 
non--standard vector type of interactions. In our analysis we use the 
helicity amplitude formalism and polarization density matrix method 
(see the first and third references in \cite{R7215}) to analyze the joint
decay distributions in this decay.  

The paper is organized as follows. In section
2, using the Hamiltonian that includes non--standard vector interactions,
the matrix element for the $\Lambda_b \rar \Lambda \ell^+ \ell^-$ is obtained.
In section 3 we calculate the different P--odd asymmetries. 
In the final section we study the sensitivity of various asymmetries to the
non--standard interactions.

\section{Matrix element for the $\Lambda_b \rar \Lambda \ell^+ \ell^-$ decay}

In this section we derive the matrix element for the $\Lambda_b \rar \Lambda
\ell^+ \ell^-$ decay which is governed by the effective Hamiltonian
describing $b \rar s \ell^+ \ell^-$ transition.
The effective Hamiltonian for the $b \rar s \ell^+ \ell^-$
transition can be written in terms of the twelve model independent
four--Fermi interactions as \cite{R7205}

\bea
\label{e7201}
{\cal M} \es \frac{G \alpha}{\sqrt{2} \pi} V_{tb}V_{ts}^\ast \Bigg\{
C_{SL} \bar s_R i \sigma_{\mu\nu} \frac{q^\nu}{q^2} b_L \bar \ell \gamma^\mu
\ell + C_{BR} \bar s_L i \sigma_{\mu\nu} \frac{q^\nu}{q^2} b_R \bar \ell
\gamma^\mu \ell + C_{LL}^{tot} \bar s_L \gamma_\mu b_L \bar \ell_L
\gamma^\mu \ell_L \nnb \\
\ar C_{LR}^{tot} \bar s_L \gamma_\mu b_L \bar \ell_R  
\gamma^\mu \ell_R + C_{RL} \bar s_R \gamma_\mu b_R \bar \ell_L
\gamma^\mu \ell_L + C_{RR} \bar s_R \gamma_\mu b_R \bar \ell_R
\gamma^\mu \ell_R \nnb \\
\ar C_{LRLR} \bar s_L b_R \bar \ell_L \ell_R +
C_{RLLR} \bar s_R b_L \bar \ell_L \ell_R +
C_{LRRL} \bar s_L b_R \bar \ell_R \ell_L +
C_{RLRL} \bar s_R b_L \bar \ell_R \ell_L \nnb \\
\ar C_T \bar s \sigma_{\mu\nu} b \bar \ell \sigma^{\mu\nu} \ell +
i C_{TE} \epsilon_{\mu\nu\alpha\beta} \bar s \sigma^{\mu\nu} b 
\bar \ell \sigma^{\alpha\beta} \ell \Bigg\}~,
\eea
where $q=P_{\Lambda_b} - P_\Lambda = p_1+p_2$ is the momentum transfer and
$C_X$ are the coefficients of the four--Fermi interactions,
$L=(1-\gamma_5)/2$ and $R=(1+\gamma_5)/2$.
The terms with coefficients $C_{SL}$ and
$C_{BR}$ describe the penguin contributions, which correspond to 
$-2 m_s C_7^{eff}$ and $-2 m_b C_7^{eff}$ in the SM, respectively. 
The next four terms in Eq. (\ref{e7201}) with coefficients
$C_{LL}^{tot},~C_{LR}^{tot},~ C_{RL}$ and $C_{RR}$
describe vector type interactions, two ($C_{LL}^{tot}$ and $C_{LR}^{tot}$)
of which contain SM contributions in the form
$C_9^{eff}-C_{10}$ and $C_9^{eff}-C_{10}$, respectively.
Thus, $C_{LL}^{tot}$ and $C_{LR}^{tot}$ can be written as 
\bea
\label{e7202}
C_{LL}^{tot} \es C_9^{eff}- C_{10} + C_{LL}~, \nnb \\
C_{LR}^{tot} \es C_9^{eff}+ C_{10} + C_{LR}~,
\eea
where $C_{LL}$ and $C_{LR}$ describe the contributions of new physics.
Additionally, Eq. (\ref{e7201}) contains four scalar type interactions 
($C_{LRLR},~C_{RLLR},~C_{LRRL}$ and $C_{RLRL}$), and two tensor type 
interactions ($C_T$ and $C_{TE}$). Note that we will neglect the
tensor type interactions throughout in this work.

The amplitude of the exclusive $\Lambda_b \rar \Lambda\ell^+ \ell^-$ decay
is obtained by calculating the matrix element of ${\cal H}_{eff}$ for the $b
\rar s \ell^+ \ell^-$ transition between initial and final
baryon states $\lla \Lambda \vel {\cal H}_{eff} \ver \Lambda_b \rra$.
It follows from Eq. (\ref{e7201}) that the matrix elements
\bea
&&\lla \Lambda \vel \bar s \gamma_\mu (1 \mp \gamma_5) b \ver \Lambda_b
\rra~,\nnb \\
&&\lla \Lambda \vel \bar s \sigma_{\mu\nu} (1 \mp \gamma_5) b \ver \Lambda_b
\rra~,\nnb
\eea
are needed in order to calculate
the $\Lambda_b \rar \Lambda\ell^+ \ell^-$ decay amplitude.

These matrix elements parametrized in terms of the form factors are 
as follows (see \cite{R7224,R7227})
\bea
\label{e7203}
\lla \Lambda \vel \bar s \gamma_\mu b \ver \Lambda_b \rra  
\es \bar u_\Lambda \Big[ f_1 \gamma_\mu + i f_2 \sigma_{\mu\nu} q^\nu + f_3  
q_\mu \Big] u_{\Lambda_b}~,\\
\label{e7204}
\lla \Lambda \vel \bar s \gamma_\mu \gamma_5 b \ver \Lambda_b \rra
\es \bar u_\Lambda \Big[ g_1 \gamma_\mu \gamma_5 + i g_2 \sigma_{\mu\nu}
\gamma_5 q^\nu + g_3 q_\mu \gamma_5\Big] u_{\Lambda_b}~, \\
\label{e7205}
\lla \Lambda \vel \bar s \sigma_{\mu\nu} b \ver \Lambda_b \rra
\es \bar u_\Lambda \Big[ f_T \sigma_{\mu\nu} - i f_T^V \ga \gamma_\mu q^\nu -
\gamma_\nu q^\mu \dr - i f_T^S \ga P_\mu q^\nu - P_\nu q^\mu \dr \Big]
u_{\Lambda_b}~,\\
\label{e7206}
\lla \Lambda \vel \bar s \sigma_{\mu\nu} \gamma_5 b \ver \Lambda_b \rra
\es \bar u_\Lambda \Big[ g_T \sigma_{\mu\nu} - i g_T^V \ga \gamma_\mu q^\nu -
\gamma_\nu q^\mu \dr - i g_T^S \ga P_\mu q^\nu - P_\nu q^\mu \dr \Big]
\gamma_5 u_{\Lambda_b}~,
\eea
where $P = p_{\Lambda_b} + p_\Lambda$ and $q= p_{\Lambda_b} - p_\Lambda$. 

The form factors of the magnetic dipole operators are defined as 
\bea
\label{e7207}
\lla \Lambda \vel \bar s i \sigma_{\mu\nu} q^\nu  b \ver \Lambda_b \rra
\es \bar u_\Lambda \Big[ f_1^T \gamma_\mu + i f_2^T \sigma_{\mu\nu} q^\nu
+ f_3^T q_\mu \Big] u_{\Lambda_b}~,\nnb \\
\lla \Lambda \vel \bar s i \sigma_{\mu\nu}\gamma_5  q^\nu  b \ver \Lambda_b \rra
\es \bar u_\Lambda \Big[ g_1^T \gamma_\mu \gamma_5 + i g_2^T \sigma_{\mu\nu}
\gamma_5 q^\nu + g_3^T q_\mu \gamma_5\Big] u_{\Lambda_b}~.
\eea

Using the identity 
\bea
\sigma_{\mu\nu}\gamma_5 = - \frac{i}{2} \epsilon_{\mu\nu\alpha\beta}
\sigma^{\alpha\beta}~,\nnb
\eea
and Eq. (\ref{e7205}), the last expression in Eq. (\ref{e7207}) can be written as
\bea
\lla \Lambda \vel \bar s i \sigma_{\mu\nu}\gamma_5  q^\nu  b \ver \Lambda_b \rra
\es \bar u_\Lambda \Big[ f_T i \sigma_{\mu\nu} \gamma_5 q^\nu \Big]
u_{\Lambda_b}~.\nnb
\eea  
Multiplying (\ref{e7205}) and (\ref{e7206}) by $i q^\nu$ and comparing with
(\ref{e7207}), one can easily obtain the following relations between the form
factors
\bea
\label{e7208}
f_2^T \es f_T + f_T^S q^2~,\crcr
f_1^T \es \Big[ f_T^V + f_T^S \ga m_{\Lambda_b} + m_\Lambda\dr \Big] 
q^2~ = - \frac{q^2}{m_{\Lambda_b} - m_\Lambda} f_3^T~,\nnb \\
g_2^T \es g_T + g_T^S q^2~,\\
g_1^T \es \Big[ g_T^V - g_T^S \ga m_{\Lambda_b} - m_\Lambda\dr \Big]
q^2 =  \frac{q^2}{m_{\Lambda_b} + m_\Lambda} g_3^T~.\nnb
\eea 

The matrix elements of scalar and pseudoscalar operators can be obtained
by multiplying both sides of Eqs.(\ref{e7203}) and (\ref{e7204}) with
$q_\mu$ and using equation of motion, as a result of which we get, 
\bea
\lla \Lambda\ve \bar{s}b\ve \Lambda_b \rra \es \frac{1}{m_b-m_s}
\bar{u}_\Lambda \Big[f_1(m_{\Lambda_b}-m_\Lambda) + f_3 q^2 \Big]
u_\Lambda~, \nnb \\
\lla \Lambda\ve \bar{s}\gamma_5 b\ve \Lambda_b \rra \es \frac{1}{m_b+m_s}
\bar{u}_\Lambda \Big[g_1(m_{\Lambda_b}+m_\Lambda) \gamma_5 - g_3 q^2 \Big]
u_\Lambda~.\nnb
\eea  

Using these definitions of the form factors, for the matrix element
of the $\Lambda_b \rar \Lambda\ell^+ \ell^-$ we get \cite{R7225,R7226}
\bea
\label{e7209}
{\cal M} \es \frac{G \alpha}{4 \sqrt{2}\pi} V_{tb}V_{ts}^\ast \frac{1}{2} \Bigg\{
\bar{\ell} \gamma_\mu (1-\gamma_5) \ell \, 
\bar{u}_\Lambda \Big[ (A_1 - D_1) \gamma_\mu (1+\gamma_5) +
(B_1 - E_1) \gamma_\mu (1-\gamma_5) \nnb \\
\ar i \sigma_{\mu\nu} q^\nu \Big( (A_2 - D_2) (1+\gamma_5) +
(B_2 - E_2) (1-\gamma_5) \Big) \Big] u_{\Lambda_b} \nnb \\
\ar \bar{\ell} \gamma_\mu (1+\gamma_5) \ell \, 
\bar{u}_\Lambda \Big[ (A_1 + D_1) \gamma_\mu (1+\gamma_5) +
(B_1 + E_1) \gamma_\mu (1-\gamma_5) \nnb \\
\ar i \sigma_{\mu\nu} q^\nu \Big( (A_2 + D_2) (1+\gamma_5) +
(B_2 + E_2) (1-\gamma_5) \Big) \nnb \\
\ar q_\mu \Big( (A_3 + D_3) (1+\gamma_5) + (B_3 + D_2) (1-\gamma_5) \Big)
\Big] u_{\Lambda_b} \nnb \\
\ar \frac{1}{2} \bar{\ell}(1-\gamma_5) \ell \, \bar{u}_\Lambda
\Big[(P_1-P_2+R_1-R_1)(1-\gamma_5) + (P_1+P_2-R_1-R_2)(1+\gamma_5)\Big]
u_{\Lambda_b} \nnb \\
\ar \frac{1}{2} \bar{\ell}(1+\gamma_5) \ell \, \bar{u}_\Lambda
\Big[(P_1-P_2+R_2-R_1)(1-\gamma_5) + (P_1+P_2+R_1+R_2)(1+\gamma_5)\Big]
u_{\Lambda_b} \Bigg\}~,\nnb \\
\eea
where
\bea
\label{e7210}
A_1 \es \frac{1}{q^2}\ga f_1^T-g_1^T \dr C_{SL} + \frac{1}{q^2}\ga
f_1^T+g_1^T \dr C_{BR} + \frac{1}{2}\ga f_1-g_1 \dr \ga C_{LL}^{tot} +
C_{LR}^{tot} \dr \nnb \\
\ar \frac{1}{2}\ga f_1+g_1 \dr \ga C_{RL} + C_{RR} \dr~,\nnb \\
A_2 \es A_1 \ga 1 \rar 2 \dr ~,\nnb \\
A_3 \es A_1 \ga 1 \rar 3 \dr ~,\nnb \\
B_1 \es A_1 \ga g_1 \rar - g_1;~g_1^T \rar - g_1^T \dr ~,\nnb \\
B_2 \es B_1 \ga 1 \rar 2 \dr ~,\nnb \\
B_3 \es B_1 \ga 1 \rar 3 \dr ~,\nnb \\
D_1 \es \frac{1}{2} \ga C_{RR} - C_{RL} \dr \ga f_1+g_1 \dr +
\frac{1}{2} \ga C_{LR}^{tot} - C_{LL}^{tot} \dr \ga f_1-g_1 \dr~,\nnb \\
D_2 \es D_1 \ga 1 \rar 2 \dr ~, \\
D_3 \es D_1 \ga 1 \rar 3 \dr ~,\nnb \\
E_1 \es D_1 \ga g_1 \rar - g_1 \dr ~,\nnb \\
E_2 \es E_1 \ga 1 \rar 2 \dr ~,\nnb \\
E_3 \es E_1 \ga 1 \rar 3 \dr ~,\nnb \\
P_1 \es \frac{1}{m_b} \Big( f_1 \ga m_{\Lambda_b} - m_\Lambda\dr + f_3 q^2
\Big) \Big( C_{LRLR} + C_{RLLR} + C_{LRRL} + C_{RLRL} \Big)~,\nnb \\
P_2 \es N_1 \ga C_{LRRL} \rar - C_{LRRL};~C_{RLRL} \rar - C_{RLRL} \dr~,\nnb \\
R_1 \es \frac{1}{m_b} \Big( g_1 \ga m_{\Lambda_b} + m_\Lambda\dr - g_3 q^2  
\Big) \Big( C_{LRLR} - C_{RLLR} + C_{LRRL} - C_{RLRL} \Big)~,\nnb \\
R_2 \es H_1 \ga C_{LRRL} \rar - C_{LRRL};~C_{RLRL} \rar - C_{RLRL} \dr~.\nnb
\eea

It follows From these expressions
that $\Lambda_b \rar\Lambda \ell^+\ell^-$ decay is described in terms of  
many form factors. It is shown in \cite{R7228} that Heavy Quark Effective
Theory (HQET) reduces
the number of independent form factors to two ($F_1$ and
$F_2$) irrelevant of the Dirac structure
of the corresponding operators, i.e., 
\bea
\label{e7211}
\lla \Lambda(p_\Lambda) \vel \bar s \Gamma b \ver \Lambda(p_{\Lambda_b})
\rra = \bar u_\Lambda \Big[F_1(q^2) + \not\!v F_2(q^2)\Big] \Gamma
u_{\Lambda_b}~,
\eea
where $\Gamma$ is an arbitrary Dirac structure and
$v^\mu=p_{\Lambda_b}^\mu/m_{\Lambda_b}$ is the four--velocity of
$\Lambda_b$. Comparing the general form of the form factors given in Eqs.
(\ref{e7204})--(\ref{e7208}) with (\ref{e7211}), one can
easily obtain the following relations among them (see also
\cite{R7224,R7225,R7227})
\bea
\label{e7212}
g_1 \es f_1 = f_2^T= g_2^T = F_1 + \sqrt{\hat{r}_\Lambda} F_2~, \nnb \\
g_2 \es f_2 = g_3 = f_3 = g_T^V = f_T^V = \frac{F_2}{m_{\Lambda_b}}~,\nnb \\
g_T^S \es f_T^S = 0 ~,\nnb \\
g_1^T \es f_1^T = \frac{F_2}{m_{\Lambda_b}} q^2~,\nnb \\
g_3^T \es \frac{F_2}{m_{\Lambda_b}} \ga m_{\Lambda_b} + m_\Lambda \dr~,\nnb \\
f_3^T \es - \frac{F_2}{m_{\Lambda_b}} \ga m_{\Lambda_b} - m_\Lambda \dr~,
\eea
where $\hat{r}_\Lambda=m_\Lambda^2/m_{\Lambda_b}^2$.

In order to obtain the helicity amplitudes for the $\Lambda_b \rar\Lambda
\ell^+\ell^-$ decay, it is convenient to regard this decay as a quasi 
two--body decay $\Lambda_b \rar\Lambda V^\ast$ followed by the leptonic
decay $V^\ast \rar  \ell^+\ell^-$, where $V^\ast$ is the off--shell $\gamma$
or $Z$ bosons. The matrix element of $\Lambda_b \rar\Lambda \ell^+\ell^-$
decay can be written in the following form:
\bea
{\cal M}_{\lambda_i}^{\lambda_\ell \bar{\lambda}_\ell} =
\sum_{\lambda_{V^\ast}} \eta_{\lambda_{V^\ast}}
L_{\lambda_{V^\ast}}^{\lambda_\ell \bar{\lambda}_\ell} \, 
H_{\lambda_{V^\ast}}^{\lambda_i}~, \nnb
\eea
where 
\bea
\label{e7213}
L_{\lambda_{V^\ast}}^{\lambda_\ell \bar{\lambda}_\ell} \es 
\varepsilon_{V^\ast}^\mu \lla \ell^-(p_\ell,\lambda_\ell) \,
\ell^+(p_\ell,\bar{\lambda}_\ell) \vel J_\mu^\ell \ver 0 \rra~,\\ 
\label{e7214}
H_{\lambda_{V^\ast}}^{\lambda_i} \es 
\ga \varepsilon_{V^\ast}^\mu \dr^\ast \lla \Lambda
(p_\Lambda,\lambda_\Lambda) \vel J_\mu^i 
\ver \Lambda_b(p_{\Lambda_b}) \rra~,
\eea
where $\varepsilon_{V^\ast}^\mu$ is the polarization vector of the virtual
intermediate vector boson. The metric tensor can be expressed in terms of
the polarization vector of the virtual vector particle
$\varepsilon_V = \varepsilon (\lambda_V)$ as
\bea
-g^{\mu\nu} = \sum_{\lambda_{V^\ast}} \eta_{\lambda_{V^\ast}}
\varepsilon_{\lambda_{V^\ast}}^\mu \varepsilon_{\lambda_{V^\ast}}^{\ast\nu}~,\nnb
\eea  
where the summation is over the helicity of the virtual vector particle
$V,~\Lambda_V = \pm 1, 0, t$ with the metric $\eta_\pm = \eta_0 = - \eta_t =
1$, where $\lambda_V = t$ is the scalar (zero) helicity component of the
virtual $V$ particle (for more details see \cite{R7228,R7229,R7230}).
The upper indices in Eqs. (\ref{e7213}) and (\ref{e7214}) correspond to the 
helicities of the leptons and the lower ones correspond to the helicity of 
the $\Lambda$ baryon. Moreover, $J_\mu^\ell$ and $J_\mu^i$ in Eqs. (\ref{e7213}) 
and (\ref{e7214}) are the leptonic and hadronic currents, respectively. 

In the calculations of the leptonic and baryonic amplitudes we will use two
different frames. The leptonic amplitude $L_{\lambda_{V^\ast}}^{\lambda_\ell
\bar{\lambda}_\ell}$ is calculated in the rest frame of the virtual vector
boson wit the z--axis chosen along the $\Lambda$ direction and the x--z
plane chosen as the virtual $V$ decay plane. The hadronic amplitude is
calculated in the rest frame of $\Lambda_b$ baryon.

Using Eqs. (\ref{e7209})--(\ref{e7214}), after lengthy calculations, we get
for the helicity amplitudes:
\bea
\label{e7215}
{\cal M}_{+1/2}^{++} \es 2 m_\ell \sin\theta \Big( H_{+1/2,+1}^{(1)} +
H_{+1/2,+1}^{(2)}\Big) + 2 m_\ell \cos\theta \Big( H_{+1/2,0}^{(1)} +
H_{+1/2,0}^{(2)}\Big) \nnb \\
\ar 2 m_\ell \Big( H_{+1/2,t}^{(1)} -
H_{+1/2,t}^{(2)}\Big) + \frac{1}{2} \sqrt{q^2} \Big[
(1+v) J_{+1/2,0}^{(1)} - (1-v) J_{+1/2,0}^{(2)} \Big] \nnb \\
\ar\frac{1}{2} \sqrt{q^2} \Big[ (1+v) J_{+1/2,t}^{(1)}
- (1-v) J_{+1/2,t}^{(2)}\Big] ~, \nnb \\
{\cal M}_{+1/2}^{+-} \es - \sqrt{q^2} (1-\cos\theta) \Big[
(1-v) H_{+1/2,+1}^{(1)} + (1+v) H_{+1/2,+1}^{(2)}\Big] -
\sqrt{q^2} \sin\theta  \Big[(1-v) H_{+1/2,0}^{(1)} \nnb \\
\ar (1+v) H_{+1/2,0}^{(2)}\Big]~, \nnb \\
{\cal M}_{+1/2}^{-+} \es \sqrt{q^2} (1+\cos\theta) \Big[
(1+v) H_{+1/2,+1}^{(1)} + (1-v) H_{+1/2,+1}^{(2)}\Big] -
\sqrt{q^2} \sin\theta  \Big[(1+v) H_{+1/2,0}^{(1)} \nnb \\
\ar (1-v) H_{+1/2,0}^{(2)}\Big]~, \nnb \\
{\cal M}_{+1/2}^{--} \es - 2 m_\ell \sin\theta \Big(
H_{+1/2,+1}^{(1)} +
H_{+1/2,+1}^{(2)}\Big) - 2 m_\ell \cos\theta \Big( H_{+1/2,0}^{(1)} +
H_{+1/2,0}^{(2)}\Big) \nnb \\
\ar 2 m_\ell \Big( H_{+1/2,t}^{(1)} -
H_{+1/2,t}^{(2)}\Big)  + \frac{1}{2} \sqrt{q^2} \Big[
(1-v) J_{+1/2,0}^{(1)} - (1+v) J_{+1/2,0}^{(2)} \Big] \nnb \\
\ar\frac{1}{2} \sqrt{q^2} \Big[ (1-v) J_{+1/2,t}^{(1)}
- (1+v) J_{+1/2,t}^{(2)}\Big]~, \nnb \\
{\cal M}_{-1/2}^{++} \es - 2 m_\ell \sin\theta \Big(
H_{-1/2,-1}^{(1)} +
H_{-1/2,-1}^{(2)}\Big) + 2 m_\ell \cos\theta \Big( H_{-1/2,0}^{(1)} +
H_{-1/2,0}^{(2)}\Big) \nnb \\
\ar 2 m_\ell \Big( H_{-1/2,t}^{(1)} -
H_{-1/2,t}^{(2)}\Big)  + \frac{1}{2} \sqrt{q^2} \Big[
(1+v) J_{-1/2,0}^{(1)} - (1-v) J_{-1/2,0}^{(2)} \Big] \nnb \\
\ar\frac{1}{2} \sqrt{q^2} \Big[ (1+v) J_{-1/2,t}^{(1)}
- (1-v) J_{-1/2,t}^{(2)}\Big]~, \nnb \\
{\cal M}_{-1/2}^{+-} \es - \sqrt{q^2} (1+\cos\theta) \Big[
(1-v) H_{-1/2,-1}^{(1)} + (1+v) H_{-1/2,-1}^{(2)}\Big] -
\sqrt{q^2} \sin\theta  \Big[(1-v) H_{-1/2,0}^{(1)} \nnb \\
\ar (1+v) H_{-1/2,0}^{(2)}\Big]~, \nnb \\
{\cal M}_{-1/2}^{-+} \es \sqrt{q^2} (1-\cos\theta) \Big[
(1+v) H_{-1/2,-1}^{(1)} + (1-v) H_{-1/2,-1}^{(2)}\Big] -
\sqrt{q^2} \sin\theta  \Big[(1+v) H_{-1/2,0}^{(1)} \nnb \\
\ar (1-v) H_{-1/2,0}^{(2)}\Big]~, \nnb \\
{\cal M}_{-1/2}^{--} \es 2 m_\ell \sin\theta \Big( H_{-1/2,-1}^{(1)} +
H_{-1/2,-1}^{(2)}\Big) - 2 m_\ell \cos\theta \Big( H_{-1/2,0}^{(1)} +
H_{-1/2,0}^{(2)}\Big) \nnb \\
\ar 2 m_\ell \Big( H_{-1/2,t}^{(1)} -
H_{-1/2,t}^{(2)}\Big)  + \frac{1}{2} \sqrt{q^2} \Big[
(1-v) J_{-1/2,0}^{(1)} - (1+v) J_{-1/2,0}^{(2)} \Big] \nnb \\
\ar\frac{1}{2} \sqrt{q^2} \Big[ (1-v) J_{-1/2,t}^{(1)}
- (1+v) J_{-1/2,t}^{(2)}\Big]~,
\eea
where
\bea      
\label{e7216}
H_{\pm 1/2,\pm1}^{(1)}   \es H_{1/2,1}^{(1)\,V}   \pm H_{1/2,1}^{(1)\,A}~,   \nnb \\
H_{\pm 1/2,\pm1}^{(2)}   \es H_{1/2,1}^{(2)\,V}   \pm H_{1/2,1}^{(2)\,A}~,   \nnb \\
H_{\pm 1/2,0}^{(1,2)}    \es H_{1/2,0}^{(1,2)\,V} \pm H_{1/2,1}^{(1,2)\,A}~, \nnb \\
H_{\pm 1/2,t}^{(1,2)}    \es H_{1/2,t}^{(1,2)\,V} \pm H_{1/2,t}^{(1,2)\,A}~,
\eea
where $\theta$ is the angle of the positron in the rest frame of the
intermediate boson with respect to its helicity axes.
Explicit expressions of the helicity amplitudes $H_{\lambda,\lambda_W}^{V,A}$
are
\bea
\label{e7217}
H_{1/2,1}^{(1)\,V} \es - \sqrt{Q_-} \Big[ F_1^V - (m_{\Lambda_b}+m_\Lambda) F_2^V
\Big]~, \nnb \\
H_{1/2,1}^{(1)\,A} \es - \sqrt{Q_+} \Big[ F_1^A + (m_{\Lambda_b}-m_\Lambda) F_2^A
\Big]~, \nnb \\
H_{1/2,1}^{(2)\,V} \es H_{1/2,1}^{(1)\,V} (F_1^V \rar F_3^V,~F_2^V \rar
F_4^V)~, \nnb \\
H_{1/2,1}^{(2)\,A} \es H_{1/2,1}^{(1)\,A} (F_1^A \rar F_3^A,~F_2^A \rar
F_4^A)~, \nnb \\
H_{1/2,0}^{(1)\,V} \es - \frac{1}{\sqrt{q^2}} \Big\{ \sqrt{Q_-} 
\Big[ (m_{\Lambda_b}+m_\Lambda) F_1^V - q^2 F_2^V \Big]
\Big\}~,\nnb \\
H_{1/2,0}^{(1)\,A} \es - \frac{1}{\sqrt{q^2}} \Big\{ \sqrt{Q_+} 
\Big[ (m_{\Lambda_b}-m_\Lambda) F_1^A + q^2 F_2^A \Big]
\Big\}~,\nnb \\
H_{1/2,0}^{(2)\,V} \es H_{1/2,0}^{(1)\,V} (F_1^V \rar F_3^V,~F_2^V \rar
F_4^V) ~,\nnb \\
H_{1/2,0}^{(2)\,A} \es H_{1/2,0}^{(1)\,A} (F_1^A \rar F_3^A,~F_2^A \rar
F_4^A) ~,\nnb \\
H_{1/2,t}^{(1)\,V} \es - \frac{1}{\sqrt{q^2}} \Big\{ \sqrt{Q_+} 
\Big[ (m_{\Lambda_b}-m_\Lambda) F_1^V + q^2 F_5^V \Big]
\Big\}~,\nnb \\
H_{1/2,t}^{(1)\,A} \es - \frac{1}{\sqrt{q^2}} \Big\{ \sqrt{Q_-} 
\Big[ (m_{\Lambda_b}+m_\Lambda) F_1^A - q^2 F_5^A \Big]
\Big\}~,\nnb \\
H_{1/2,t}^{(2)\,V} \es H_{1/2,t}^{(1)\,V} (F_1^V \rar F_3^V,~F_5^V \rar 
F_6^V) ~,\nnb \\
H_{1/2,t}^{(2)\,A} \es H_{1/2,t}^{(1)\,A} (F_1^A \rar F_3^A,~F_5^A \rar
F_6^A) ~, \nnb \\
J_{+1/2,0}^{(1)} \es J_{+1/2,t}^{(1)} = \sqrt{Q_+} (P_1 - P_2) - \sqrt{Q_-} (R_1 - R_2) ~, \nnb \\
J_{+1/2,0}^{(2)} \es J_{+1/2,t}^{(2)} = J_{+1/2,0}^{(1)} (P_2 \rar - P_2,~R_2 \rar - R_2)~,\nnb \\
J_{-1/2,0}^{(1)} \es J_{-1/2,t}^{(1)} = J_{+1/2,0}^{(1)} (\sqrt{Q_-} \rar - \sqrt{Q_-})~,\nnb \\
J_{-1/2,0}^{(2)} \es J_{-1/2,t}^{(2)} = J_{+1/2,0}^{(2)} (\sqrt{Q_-} \rar - \sqrt{Q_-})~, 
\eea
where
\bea
Q_+ \es (m_{\Lambda_b}+m_\Lambda)^2 - q^2~,\nnb \\
Q_- \es (m_{\Lambda_b}-m_\Lambda)^2 - q^2~,\nnb
\eea 
and
\bea
\label{e7218}
F_1^V \es A_1-D_1+B_1-E_1~,\nnb \\
F_1^A \es A_1-D_1-B_1+E_1~,\nnb \\
F_2^V \es F_1^V (1\rar 2)~,\nnb \\
F_2^A \es F_1^A (1\rar 2)~,\nnb \\
F_3^V \es A_1+D_1+B_1+E_1~,\nnb \\
F_3^A \es A_1+D_1-B_1-E_1~,\nnb \\
F_4^V \es F_3^V (1\rar 2)~,\nnb \\
F_4^A \es F_3^A (1\rar 2)~,\nnb \\
F_5^V \es F_1^V (1\rar 3)~,\nnb \\
F_5^A \es F_1^A (1\rar 3)~,\nnb \\
F_6^V \es F_4^V (2\rar 3)~,\nnb \\
F_6^A \es F_4^A (2\rar 3)~.
\eea
The remaining helicity amplitudes can be obtained from the parity relations
\bea
\label{e7219}
H_{-\lambda,-\lambda_W}^{V,(A)} = +(-) H_{\lambda,\lambda_W}^{V,(A)}~.
\eea
The square of the matrix element for the $\Lambda_b \rar\Lambda \ell^+\ell^-$ 
decay is given as
\bea
\label{e7220}
\vel {\cal M} \ver^2 \es \vel {\cal M}_{+1/2}^{++} \ver^2 +
\vel {\cal M}_{+1/2}^{+-} \ver^2 + \vel {\cal M}_{+1/2}^{-+} \ver^2 + 
\vel {\cal M}_{+1/2}^{--} \ver^2 \nnb \\
\ar \vel {\cal M}_{-1/2}^{++} \ver^2 +
\vel {\cal M}_{-1/2}^{+-} \ver^2 + \vel {\cal M}_{-1/2}^{-+} \ver^2
+ \vel {\cal M}_{-1/2}^{--} \ver^2~.
\eea
Following the standard methods used in literature (see the third
reference in \cite{R7215}), the normalized joint angular decay
distribution for the two cascade decay
\bea
\Lambda_b^{1/2 ^+} \rar \Lambda^{1/2 ^+} \Big( \rar a(1/2 ^+) +
b(0^-)\Big)
+ V (\rar \ell^+ \ell^-)~,\nnb 
\eea
\bea
\label{e7221}
\lefteqn{
\frac{\ds d \Gamma}{\ds dq^2 d\!\cos\theta \,d\!\cos\theta_\Lambda} =
\vel \frac{G \alpha}{4 \sqrt{2} \pi} V_{tb} V_{ts}^\ast \frac{1}{2} \ver^2
\frac{\sqrt{\lambda(m_{\Lambda_b}^2,m_\Lambda^2,q^2)}
\sqrt{\lambda(m_\Lambda^2,m_a^2,m_b^2)}}{1024 \pi^3 m_{\Lambda_b}^3 m_\Lambda^2}
v {\cal B}(\Lambda \rar a + b) }\nnb \\ 
&&\Bigg\{
(1+\alpha_\Lambda \cos \theta_\Lambda) \Bigg[ \Big(8 m_\ell^2 \sin^2\theta 
\vel A_{+1/2,+1} \ver^2 + (1-\cos\theta)^2 q^2 
\vel A_{+1/2,+1}- v B_{+1/2,+1} \ver^2 \nnb \\
&&+ (1+\cos\theta)^2 q^2 
\vel A_{+1/2,+1}+ v B_{+1/2,+1} \ver^2 \Big) +
 8 m_\ell^2 \cos^2\theta \vel A_{+1/2,0} \ver^2 +
8 m_\ell^2 \vel B_{+1/2,t} \ver^2 \nnb \\
&& + \sin^2\theta q^2 \Big( 2 \vel A_{+1/2,0} \ver^2 + 2 v^2 \vel B_{+1/2,0} \ver^2
\Big) \nnb \\
&&-  4 m_\ell \sqrt{q^2} \Big( \mbox{\rm Re}[B_{+1/2,t}(D_{+1/2,t}^\ast +
D_{+1/2,0}^\ast)] + v \cos\theta \mbox{\rm Re}[A_{+1/2,0}(C_{+1/2,t}^\ast
+C_{+1/2,0}^\ast)]\Big) \nnb \\
&&+ \frac{q^2}{2} \Big( \vel D_{+1/2,t} + D_{+1/2,0} \ver^2 + 
v^2 \vel C_{+1/2,t} + C_{+1/2,0} \ver^2 \Big) 
\Bigg] \nnb \\
&& + (1-\alpha_\Lambda \cos \theta_\Lambda) \Bigg[ \Big(8 m_\ell^2 \sin^2\theta 
\vel A_{-1/2,-1} \ver^2 + (1+\cos\theta) q^2 
\vel A_{-1/2,-1}- v B_{-1/2,-1} \ver^2 \nnb \\
&&+ (1-\cos\theta)^2 q^2 
\vel A_{-1/2,-1}+ v B_{-1/2,-1} \ver^2 \Big) +
8 m_\ell^2 \cos^2\theta \vel A_{-1/2,0} \ver^2 +
8 m_\ell^2 \vel B_{-1/2,t} \ver^2 \nnb \\
&& + \sin^2\theta q^2 \Big( 2 \vel A_{-1/2,0} \ver^2 + 2 v^2 \vel B_{-1/2,0} \ver^2
\Big) \nnb \\
&&- 4 m_\ell \sqrt{q^2} \Big( \mbox{\rm Re}[B_{-1/2,t}(D_{-1/2,t}^\ast +
D_{-1/2,0}^\ast)] + v \cos\theta \mbox{\rm Re}[A_{-1/2,0}(C_{-1/2,t}^\ast
+C_{-1/2,0}^\ast)]\Big) \nnb \\
&&+ \frac{q^2}{2} \Big( \vel D_{-1/2,t} + D_{-1/2,0} \ver^2 + 
v^2 \vel C_{-1/2,t} + C_{-1/2,0} \ver^2 \Big)
\Bigg]\Bigg\}~.
\eea

In Eq. (\ref{e7221}) we induce the following definitions:
\bea
\label{e7222}
H_{\lambda_i,\lambda_W}^{(1)} + H_{\lambda_i,\lambda_W}^{(2)} \es 
A_{\lambda_i,\lambda_W}~,\nnb \\
H_{\lambda_i,\lambda_W}^{(1)} - H_{\lambda_i,\lambda_W}^{(2)} \es 
B_{\lambda_i,\lambda_W}~,\nnb \\
J_{\lambda_i,\lambda_W}^{(1)} + J_{\lambda_i,\lambda_W}^{(2)} \es 
C_{\lambda_i,\lambda_W}~,\nnb \\
J_{\lambda_i,\lambda_W}^{(1)} - J_{\lambda_i,\lambda_W}^{(2)} \es 
D_{\lambda_i,\lambda_W}~.
\eea

Note that in deriving Eq. (\ref{e7222}), we perform integration over the
azimuthal angle $\varphi$ between the planes of the two decays
$\Lambda \rar a+b$ and $V \rar \ell^+ \ell^-$.

It is well known
that heavy quarks $b(c)$ resulting from $Z$ decay are polarized. It is shown
in \cite{R7231,R7232} that a sizeable fraction of the $b$ quark polarization
retained in fragmentation of heavy quarks to heavy baryons. Therefore, an
additional set of polarization observables can be obtained if the
polarization of the heavy $\Lambda_b$ baryon is taken into account.

In order to take polarization of the $\Lambda_b$ baryon into consideration    
we will use the density matrix method. The spin density matrix of $\Lambda$
baryon is
\bea
\label{e7223}
\rho = \frac{1}{2}
\left(
\begin{array}{cc}
1+{\cal P}\cos\theta_\Lambda^S & {\cal P}\cos\theta_\Lambda^S \\ \\
{\cal P}\cos\theta_\Lambda^S   & 1-{\cal P}\cos\theta_\Lambda^S
\end{array}
\right)~,
\eea
where ${\cal P}$ is the polarization of $\Lambda_b$, and $\theta_\Lambda^S$
is the angle that the polarization of $\Lambda_b$ makes with the momentum of
$\Lambda$, in the rest frame of $\Lambda_b$.  

The four--fold decay distribution can easily be obtained from Eq.
(\ref{e7221}). Obviously, there appears on the left--hand side of Eq.
(\ref{e7221}) the distribution over $\theta_\Lambda^S$, i.e., 
$d/d\cos\theta_\Lambda^S$. Hence the right--hand side of the same equation
can be modified as follows:
\bea
\label{e7224}
&&\vel 1/2,1 \ver^2 \rar (1-{\cal P}\cos\theta_\Lambda^S) \vel 1/2,1
\ver^2~,\nnb \\
&&\vel 1/2,t \ver^2,~\vel 1/2,0 \ver^2,~ (1/2,0) (1/2,t)^\ast 
\rar (1+{\cal P}\cos\theta_\Lambda^S)
\Big\{ \vel 1/2,0 \ver^2,\vel 1/2,t \ver^2,\nnb \\
&&(1/2,0) (1/2,t)^\ast \Big\}~, \nnb \\ 
&&(1/2,1) (1/2,t)^\ast,~(1/2,1) (1/2,0)^\ast \rar
{\cal P}\sin\theta_\Lambda^S \Big\{(1/2,1) (1/2,0)^\ast,~(1/2,1)
(1/2,t)^\ast \Big\}~, \nnb \\
&&\vel -1/2,-1 \ver^2 \rar (1+{\cal P}\cos\theta_\Lambda^S) \vel -1/2,-1 \ver^2 
~,\nnb \\
&&\vel - 1/2,t \ver^2,~\vel -1/2,0 \ver^2,~(-1/2,0) (-1/2,t)^\ast 
\rar (1-{\cal P}\cos\theta_\Lambda^S)
\Big\{ \vel -1/2,0 \ver^2,~\vel -1/2,t \ver^2,\nnb \\
&&(-1/2,0) (-1/2,t)^\ast\Big\}~, 
\nnb \\
&&(-1/2,-1) (-1/2,-t)^\ast,~(-1/2,-1) (-1/2,0)^\ast \rar
{\cal P}\sin\theta_\Lambda^S \Big\{(-1/2,-1) (-1/2,t)^\ast,\nnb \\
&&(-1/2,-1)(-1/2,0)^\ast \Big\}~.
\eea

From the expressions for the four--fold angular distribution we may define
the following forward--backward asymmetries:
\bea
\label{e7225}
{\cal A}_\theta^{FB} \es \frac{\ds
\left[ \int_0^{+1} d\!\cos\theta - \int_{-1}^0 d\!\cos\theta \right]
\int_{-1}^{+1} d\!\cos\theta_\Lambda \int_{-1}^{+1}
d\!\cos\theta_\Lambda^S
\frac{d\Gamma}{dq^2\,d\!\cos\theta\,d\!\cos\theta_\Lambda\,
d\!\cos\theta_\Lambda^S } }
{\ds
\int_{-1}^{+1} d\!\cos\theta\,\int_{-1}^{+1} d\!\cos\theta_\Lambda
\int_{-1}^{+1} d\!\cos\theta_\Lambda^S
\frac{d\Gamma}{dq^2\,d\!\cos\theta\,d\!\cos\theta_\Lambda\,
d\!\cos\theta_\Lambda^S
} }~,\nnb \\ \\ 
\label{e7226}
{\cal A}_{\theta_\Lambda^S}^{FB} \es \frac{\ds
\left[ \int_0^{+1} d\!\cos\theta_\Lambda^S - \int_{-1}^0
d\!\cos\theta_\Lambda^S \right]
\int_{-1}^{+1} d\!\cos\theta \int_{-1}^{+1}
d\!\cos\theta_\Lambda
\frac{d\Gamma}{dq^2\,d\!\cos\theta\,d\!\cos\theta_\Lambda\,
d\!\cos\theta_\Lambda^S } }
{\ds
\int_{-1}^{+1} d\!\cos\theta\,\int_{-1}^{+1} d\!\cos\theta_\Lambda
\int_{-1}^{+1} d\!\cos\theta_\Lambda^S
\frac{d\Gamma}{dq^2\,d\!\cos\theta\,d\!\cos\theta_\Lambda\,
d\!\cos\theta_\Lambda^S
} }~,\nnb \\ \\
\label{e7227}
{\cal A}_{\theta_\Lambda}^{FB} \es \frac{\ds
\left[ \int_0^{+1} d\!\cos\theta_\Lambda - \int_{-1}^0
d\!\cos\theta_\Lambda \right]
\int_{-1}^{+1} d\!\cos\theta \int_{-1}^{+1}
d\!\cos\theta_\Lambda^S
\frac{d\Gamma}{dq^2\,d\!\cos\theta\,d\!\cos\theta_\Lambda\,
d\!\cos\theta_\Lambda^S } }
{\ds
\int_{-1}^{+1} d\!\cos\theta\,\int_{-1}^{+1} d\!\cos\theta_\Lambda
\int_{-1}^{+1} d\!\cos\theta_\Lambda^S
\frac{d\Gamma}{dq^2\,d\!\cos\theta\,d\!\cos\theta_\Lambda\,
d\!\cos\theta_\Lambda^S
} }~.\nnb \\ 
\eea
Performing relevant integrations in Eqs. (\ref{e7225})--(\ref{e7227}), we
get:
\bea
\label{e7228}
{\cal A}_\theta^{FB} \es \frac{16 v \sqrt{q^2} }{\Delta} 
\Big\{2 \sqrt{q^2}  \mbox{\rm Re}[ 
A_{+1/2,+1} B_{+1/2,+1} - A_{-1/2,-1} B_{-1/2,-1} ] \nnb \\
\ek m_\ell \mbox{\rm Re}[ A_{-1/2,0} (C_{-1/2,t}^\ast +
C_{-1/2,0}^\ast) + A_{+1/2,0} (C_{+1/2,t}^\ast + C_{+1/2,0}^\ast)]
\Big\}~, \\ \nnb \\
\label{e7229}
{\cal A}_{\theta_\Lambda^S}^{FB} \es \frac{2}{3 \Delta} \Big\{
- 32 m_\ell^2 \vel A_{+1/2,+1} \ver^2 - 16 q^2
\Big(\vel A_{+1/2,+1} \ver^2 + v^2 \vel B_{+1/2,+1} \ver^2
\Big) \nnb \\
\ar 16 m_\ell^2 \vel A_{+1/2,0} \ver^2
+ 8 q^2 \Big(\vel A_{+1/2,0}
\ver^2 + v^2 \vel B_{+1/2,0} \ver^2 \Big) \nnb \\
\ar 32 m_\ell^2 \vel
A_{-1/2,-1} \ver^2 + 16 q^2
\Big(\vel A_{-1/2,-1} \ver^2 + v^2 \vel B_{-1/2,-1} \ver^2 \Big) \nnb \\
\ek 16 m_\ell^2 \vel A_{-1/2,0} \ver^2 -
8 q^2 \big(\vel A_{-1/2,0} \ver^2 + v^2 \vel B_{-1/2,0}
\ver^2\Big) \nnb \\
\ar 48 m_\ell^2 \vel B_{+1/2,t} \ver^2 -
48 m_\ell^2 \vel B_{-1/2,t} \ver^2 \nnb \\
\ek 3 q^2 \vel D_{-1/2,t} \ver^2 +
24 m_\ell \sqrt{q^2}
\mbox{\rm Re}[B_{-1/2,t}(D_{-1/2,t}^\ast + D_{-1/2,0}^\ast)
- B_{+1/2,t}(D_{+1/2,t}^\ast + D_{+1/2,0}^\ast)] \nnb \\
\ar 3 q^2 \Big( \vel D_{+1/2,t} \ver^2 + \vel D_{+1/2,0} \ver^2 
- \vel D_{-1/2,0} \ver^2 + 
2 \mbox{\rm Re}[D_{+1/2,0} D_{+1/2,t}^\ast - D_{-1/2,0} D_{-1/2,t}^\ast] \Big) \nnb \\
\ar 3 q^2 v^2 \Big( \vel C_{+1/2,t} \ver^2 - \vel C_{-1/2,t} \ver^2 
- \vel C_{-1/2,0} \ver^2 + 
2 \mbox{\rm Re}[C_{+1/2,t} C_{+1/2,0}^\ast - C_{-1/2,t} C_{-1/2,0}^\ast] \Big)
\Big\} ~, \nnb \\ \\
\label{e7230}
{\cal A}_{\theta_\Lambda}^{FB} \es \frac{2}{3 \Delta} \Big\{
32 m_\ell^2 \vel A_{+1/2,+1} \ver^2 + 16 q^2
\Big(\vel A_{+1/2,+1} \ver^2 + v^2 \vel B_{+1/2,+1} \ver^2
\Big) \nnb \\
\ar 16 m_\ell^2 \vel A_{+1/2,0} \ver^2
+ 8 q^2 \Big(\vel A_{+1/2,0}
\ver^2 + v^2 \vel B_{+1/2,0} \ver^2 \Big) \nnb \\
\ek 32 m_\ell^2 \vel
A_{-1/2,-1} \ver^2 - 16 q^2
\Big(\vel A_{-1/2,-1} \ver^2 + v^2 \vel B_{-1/2,-1} \ver^2 \Big) \nnb \\
\ek 16 m_\ell^2 \vel A_{-1/2,0} \ver^2 -
8 q^2 \big(\vel A_{-1/2,0} \ver^2 + v^2 \vel B_{-1/2,0}
\ver^2\Big) \nnb \\
\ar 48 m_\ell^2 \vel B_{+1/2,t} \ver^2 -
48 m_\ell^2 \vel B_{-1/2,t} \ver^2 \nnb \\
\ek 3 q^2 \vel D_{-1/2,t} \ver^2 + 
24 m_\ell \sqrt{q^2} \mbox{\rm Re}[ B_{-1/2,t} (D_{-1/2,t}^\ast +
D_{-1/2,0}^\ast) - B_{+1/2,t} (D_{+1/2,t}^\ast + D_{+1/2,0}^\ast)] \nnb \\
\ar 3 q^2 \Big( \vel D_{+1/2,t} \ver^2 + \vel D_{+1/2,0} \ver^2 -
\vel D_{-1/2,0} \ver^2 - 2 \mbox{\rm Re}[ D_{-1/2,t} D_{-1/2,0}^\ast -
D_{+1/2,t} D_{+1/2,0}^\ast] \Big) \nnb \\
\ar 3 q^2 v^2 \Big( \vel C_{+1/2,t} \ver^2 - \vel C_{-1/2,t}\ver^2 - 
\vel C_{-1/2,0} \ver^2 - 2 \mbox{\rm Re}[ C_{-1/2,t} C_{-1/2,0}^\ast -
C_{+1/2,t} C_{+1/2,0}^\ast]\Big) \Big\}~, \nnb \\   
\eea
where
\bea
\label{e7231}
\Delta \es \frac{4}{3} \Big\{ 32 m_\ell^2 \vel A_{+1/2,+1} \ver^2 + 16 q^2
\Big(\vel A_{+1/2,+1} \ver^2 + v^2 \vel B_{+1/2,+1} \ver^2
\Big) \nnb \\
\ar 16 m_\ell^2 \vel A_{+1/2,0} \ver^2
+ 8 q^2 \Big(\vel A_{+1/2,0}
\ver^2 + v^2 \vel B_{+1/2,0} \ver^2 \Big) \nnb \\
\ar 32 m_\ell^2 \vel
A_{-1/2,-1} \ver^2 + 16 q^2
\Big(\vel A_{-1/2,-1} \ver^2 + v^2 \vel B_{-1/2,-1} \ver^2 \Big) \nnb \\
\ar 16 m_\ell^2 \vel A_{-1/2,0} \ver^2 +
8 q^2 \big(\vel A_{-1/2,0} \ver^2 + v^2 \vel B_{-1/2,0}
\ver^2\Big) \nnb \\
\ar 48 m_\ell^2 \vel B_{+1/2,t} \ver^2 +
48 m_\ell^2 \vel B_{-1/2,t} \ver^2 \nnb \\
\ar 3  q^2 \Big( \vel D_{-1/2,t} \ver^2 + \vel D_{-1/2,0} \ver^2 + 
\vel D_{+1/2,t} \ver^2 + \vel D_{+1/2,0} \ver^2 +
2 \mbox{\rm Re}[D_{-1/2,t} D_{-1/2,0}^\ast + D_{+1/2,t} D_{+1/2,0}^\ast]
\Big) \nnb \\ 
\ek 24 m_\ell \sqrt{q^2} \mbox{\rm Re}[B_{-1/2,t} (D_{-1/2,t}^\ast +
D_{-1/2,0}^\ast) + B_{+1/2,t} (D_{+1/2,t}^\ast + D_{+1/2,0}^\ast)] \nnb \\
\ar 3 q^2 v^2 \Big( \vel C_{-1/2,t} \ver^2 + \vel C_{-1/2,0} \ver^2 + 
\vel C_{+1/2,t} \ver^2 + \vel C_{+1/2,0} \ver^2 +
2 \mbox{\rm Re}[C_{-1/2,t} C_{-1/2,0}^\ast + C_{+1/2,t} C_{+1/2,0}^\ast]
\Big) \Big\} ~. \nnb \\   
\eea

\section{Numerical analysis}

In this section we will study the sensitivity of the P--odd asymmetries on
the new Wilson coefficients.
The values of the input parameters we use in our
calculations are: $\vel V_{tb} V_{ts}^\ast \ver = 0.0385$, 
$m_\tau = 1.77~GeV$, $m_\mu = 0.106~GeV$,
$m_b = 4.8~GeV$, and we neglect the mass of the strange quark. 
In further numerical analysis, the values of the new Wilson 
coefficients which describe new physics beyond the SM, are needed. 
In numerical calculations we will vary all new Wilson coefficients in the 
range $- \vel C_{10}^{SM} \ver \le C_X \le \vel C_{10}^{SM} \ver$. The
experimental results on the branching ratio of the 
$B \rar K^\ast \ell^+ \ell^-$ decay \cite{R7212,R7213} and the bound on the
branching ratio of $B \rar \mu^+ \mu^-$ \cite{R7235} suggest that 
this is the right order of
magnitude for the vector and scalar interaction coefficients.
For the values of the Wilson coefficients in the SM 
we use: $C_7^{SM} = -0.313$, $C_9^{SM} = 4.344$ and $C_{10}^{SM} = -4.669$.
The magnitude of $C_7^{eff}$ is quite well constrained from $b \rar s
\gamma$ decay, and hence well established. Moreover, we will fix the values 
of the Wilson coefficients,
i.e., $C_{BR}$ and $C_{SL}$ are both related to $C_7^{eff}$ as follows:
$C_{BR}=-2 m_b C_7^{eff}$ and $C_{SL}=-2 m_s C_7^{eff}$.  
As far as the Wilson coefficient
$C_9^{SM}$ is considered, we take into account the short, as well as the
long distance contributions coming from the production of
$\bar{c}c$ pair at intermediate states. It is well known that the form
factors are the main and the most important 
input parameters necessary in
the numerical calculations. The calculation of the form factors of $\Lambda_b
\rar \Lambda$ transition does not exist at present.
But, we can use the results from QCD sum rules
in corporation with HQET \cite{R7228,R7233}. We noted earlier that,
HQET allows us to establish relations among the form factors and reduces
the number of independent form factors into two. 
In \cite{R7228,R7233}, the $q^2$ dependence of these form factors
are given as follows
\bea
F(\hat{s}) = \frac{F(0)}{\ds 1-a_F \hat{s} + b_F \hat{s}^2}~. \nnb
\eea
The values of the parameters $F(0),~a_F$ and $b_F$ are given in table 1.
\begin{table}[h]    
\renewcommand{\arraystretch}{1.5}
\addtolength{\arraycolsep}{3pt}
$$
\begin{array}{|l|ccc|}  
\hline
& F(0) & a_F & b_F \\ \hline
F_1 &
\phantom{-}0.462 & -0.0182 & -0.000176 \\
F_2 &
-0.077 & -0.0685 &\phantom{-}0.00146 \\ \hline
\end{array}
$$
\caption{Form factors for $\Lambda_b \rar \Lambda \ell^+ \ell^-$
decay in a three parameter fit.}
\renewcommand{\arraystretch}{1}
\addtolength{\arraycolsep}{-3pt}
\end{table}  

In order to have an idea about the sensitivity of our results to the
specific parametrization of the two form factors predicted by the QCD sum
rules in corporation with the HQET, we also have used another
parametrization of the form factors based on the pole model and compared the
results of both models. The dipole form of the form factors predicted by the
pole model are given as:
\bea
F_{1,2}(E_\Lambda) = N_{1,2} \left( \frac{\Lambda_{QCD}}{\Lambda_{QCD}+
E_\Lambda} \right)^2~,\nnb
\eea
where
\bea
E_\Lambda = \frac{m_{\Lambda_b}^2 - m_\Lambda^2 - q^2}
{2 m_{\Lambda_b}}~,\nnb
\eea
and $\Lambda_{QCD} = 0.2$, $\vel N_1\ver = 52.32$ and $\vel N_1\ver \simeq
-0.25 N_1$ \cite{R7234}.

It is well known that, in addition to the short distance contributions
$C_{9}$ receives long distance contributions coming from the production
of the real $\bar{c}c$ state that can be written as
\bea
\label{e7235} 
C_9^{eff}(m_b,\hat s) = C_9(m_b)\left[1 + \frac{\alpha_s(\mu)}{\pi} \omega
(\hat s) \right] + Y(\hat s)~,
\eea
where $C_9(m_b)=4.344$ and 
\bea
Y(\hat s) = Y^{per}(\hat s) + Y_{LD}~.\nnb
\eea
Here 
\bea
Y^{per}(\hat s) &=& g \ga \hat m_c,\hat s \dr
C^{(0)}
- \frac{1}{2} g \ga 1,\hat s \dr
\left[4 C_3 +4 C_4 + 3 C_5 + C_6 \right] \nnb \\
&-& \frac{1}{2} g \ga 0,\hat s \dr
\left[ C_3 + 3  C_4 \right]
+ \frac{2}{9} \left[ 3 C_3 + C_4 + 3 C_5 + C_6 \right] \nnb \\
&-& \lambda_u 
\left[ 3 C_1 + C_2 \right] \left[ g \ga 0,\hat s \dr -
g \ga \hat m_c,\hat s \dr \right]~,\nnb
\eea
where 
\bea
C^{(0)} &=& 3 C_1 + C_2 + 3 C_3 + C_4 + 3 C_5 + C_6~,\nnb \\ \nnb \\
\lambda_u &=& \frac{V_{ub} V_{ud}^\ast}{V_{tb} V_{td}^\ast}~, \nnb
\eea
and the loop function $g \ga m_q, s \dr$ stands for the loops
of quarks with mass $m_{q}$ at the dilepton invariant mass $s$.
This function develops absorptive parts for dilepton energies  
$s= 4 m_q^{2}$:
\bea
\lefteqn{
g \ga \hat m_q,\hat s \dr = - \frac{8}{9} \ln \hat m_q +
\frac{8}{27} + \frac{4}{9} y_q -
\frac{2}{9} \ga 2 + y_q \dr \sqrt{\vel 1 - y_q \ver}} \nnb \\
&&\times \Bigg[ \Theta(1 - y_q)
\ga \ln \frac{1  + \sqrt{1 - y_q}}{1  -  \sqrt{1 - y_q}} - i \pi \dr
+ \Theta(y_q - 1) \, 2 \, \arctan \frac{1}{\sqrt{y_q - 1}} \Bigg], \nnb
\eea
where  $\hat m_q= m_{q}/m_{b}$ and $y_q=4 \hat m_q^2/\hat s$
(see \cite{R7236,R7237}). 
The long distance contributions are embedded into $Y_{LD}$ whose expression
is given as 
\bea
Y_{LD}(\hat s) &=& \frac{3}{\alpha^2}
\Bigg[ - \frac{V_{cf}^* V_{cb}}{V_{tf}^* V_{tb}} \,C^{(0)} -
\frac{V_{uf}^* V_{ub}}{V_{tf}^* V_{tb}}
\ga 3 C_3 + C_4 + 3 C_5 + C_6 \dr \Bigg] \nnb \\
&\times& \sum_{V_i = \psi \ga 1 s \dr, \cdots, \psi \ga 6 s \dr}
\ds{\frac{ \pi \kappa_{i} \Gamma \ga V_i \rar \ell^+ \ell^- \dr M_{V_i} }
{\ga M_{V_i}^2 - \hat s m_b^2 - i M_{V_i} \Gamma_{V_i} \dr }}~, \nnb
\eea
where $\kappa_i$ are the Fudge factors (see for example \cite{R7238}).

From the explicit expressions of the asymmetry parameters we see that they
depend on the new Wilson coefficients and $q^2$. Therefore there might
appear some difficulties in studying the dependence of the physical
quantities on both variables in the experiments. In order to avoid this
difficulty we perform our analysis at fixed values of $C_X$.

In Fig. (1), the dependence of the P--odd asymmetry ${\cal A}_\theta^{FB}$
on $q^2$ for the $\Lambda_b \rar \Lambda \mu^+ \mu^-$ decay is presented
at five different values of $C_{LL}$.
From this figure we see that, outside the resonance regions, the 
zero--position of ${\cal A}_\theta^{FB}$ is shifted compared to that of 
the SM result and this behavior of ${\cal A}_\theta^{FB}$ is quite similar 
to the one determined by the coefficient $C_{LR}$. The zero--positions of
${\cal A}_\theta^{FB}$ occur at the values $q^2 < 5~GeV^2$, and therefore  
the zero of ${\cal A}_\theta^{FB}$ is sensitive only to the short distance
contributions of the new Wilson coefficients and is free of the long
distance effects. When new Wilson coefficients are negative (positive), the
zero--position of ${\cal A}_\theta^{FB}$ in the SM shifts to right (left).
Further analysis shows that the zero--position of ${\cal A}_\theta^{FB}$ is
practically independent of the Wilson coefficients $C_{RR}$ and $C_{RL}$ and
coincides with that of the SM result. Moreover, for the $\Lambda_b \rar
\Lambda \mu^+ \mu^-$ decay, ${\cal A}_\theta^{FB}$ seems to be 
insensitive to the presence of any scalar type interactions in the allowed 
region of the new Wilson coefficients.   

Depicted in Fig. (2) is the effect of the Wilson coefficient
$C_{LL}$ on the dependence of ${\cal A}_\theta^{FB}$ on $q^2$
for the $\Lambda_b \rar \Lambda \tau^+ \tau^-$ decay. We observe from this
figure that, far from the resonance region, the zero--position of 
${\cal A}_\theta^{FB}$ for this decay channel is realized only for
$C_{LL}=-4$. Similarly, another zero--position occurs at
$C_{LR}=-4$. Therefore, the analysis of the zero--position, as well as    
determination of ${\cal A}_\theta^{FB}$
between the resonance regions, can serve as a good test for establishing new
physics beyond the SM. In the presence of the Wilson coefficients $C_{LL}$, 
$C_{LR}$, $C_{RL}$ and $C_{RR}$,  the absolute value of ${\cal A}_\theta^{FB}$ 
in the SM differs, approximately, two times compared to its absolute value
in the new physics beyond the SM, between the resonance regions. 
Therefore, measurement of the value of ${\cal A}_\theta^{FB}$ in experiments
can give useful information about the new physics. It is further observed that
${\cal A}_\theta^{FB}$ for the $\Lambda_b \rar \Lambda \tau^+ \tau^-$ decay
is very sensitive to the existence of the scalar interaction with the
coefficient $C_{LRRL}$, while independent of all other scalar interactions 
(see Fig. (3)). Therefore measurement of ${\cal A}_\theta^{FB}(\Lambda_b
\rar \Lambda \tau^+ \tau^-)$ can be quite informative for establishing   
the new scalar interactions.

In Figs, (4) and (5), we present the dependence of
${\cal A}_{\theta_\Lambda}^{FB}$ on $q^2$ at five fixed values of the 
Wilson coefficients $C_{RR}$ and $C_{RL}$, respectively, for the
$\Lambda_b \rar \Lambda \mu^+ \mu^-$ decay. From these figures we see that,
up to $q^2 = 18~GeV^2$, the magnitude of ${\cal A}_{\theta_\Lambda}^{FB}$
decreases at all values of the Wilson coefficients. Contrary to the 
${\cal A}_\theta^{FB}$ case, where ${\cal A}_\theta^{FB}$ exhibits strong
dependence on $C_{LL}$ and $C_{LR}$, 
${\cal A}_{\theta_\Lambda}^{FB}$ is practically insensitive to these 
coefficients.

In Fig. (6), we present the dependence of ${\cal A}_{\theta_\Lambda}^{FB}$
on $q^2$ for the $\Lambda_b \rar \Lambda \tau^+ \tau^-$ decay, at several 
fixed values of the Wilson coefficients $C_{RL}$. It should be noted that,
${\cal A}_{\theta_\Lambda}^{FB}$ shows practically similar behavior on
$C_{RR}$, and for this reason we present only the result for $C_{RL}$.
We observe that, the zero--position is absent for the $\Lambda_b \rar
\Lambda \tau^+ \tau^-$ decay. However, a measurement of the magnitude of
${\cal A}_{\theta_\Lambda}^{FB}$ can give conclusive information about the
existence of the new physics. Similar to the $\Lambda_b \rar \Lambda \mu^+
\mu^-$ case, ${\cal A}_{\theta_\Lambda}^{FB}$ is weakly dependent on the
Wilson coefficients $C_{LL}$ and $C_{LR}$. It is observed that, ${\cal
A}_{\theta_\Lambda}^{FB}$ for the $\Lambda_b \rar \Lambda \tau^+ \tau^-$
decay is rather sensitive to the scalar interactions with the coefficients
$C_{RLRL}$ and $C_{RLLR}$, while it is independent of the remaining scalar
interaction coefficients. Close to the end of the allowed region
($q^2>18~GeV^2$), ${\cal A}_{\theta_\Lambda}^{FB}(\Lambda_b \rar \Lambda
\tau^+ \tau^-)$ shows considerable departure from the SM result (see Fig.
(7)).  
 
Our analysis shows that the zero--position of 
${\cal A}_{\theta_\Lambda^S}^{FB}$ for the $\Lambda_b \rar \Lambda \mu^+
\mu^-$ case is practically independent of the new vector interaction
with coefficients $C_{LR}$, $C_{RL}$ and $C_{RR}$, and only in the presence
of the coefficient $C_{LL}$ it shifts to the right (left) compared to the SM
prediction, at its negative (positive) values. Moreover, the value of ${\cal
A}_{\theta_\Lambda^S}^{FB}$ shows considerable departure from the SM values
for the coefficients $C_{LR}$, $C_{RL}$ and $C_{RR}$ in the region
$2~GeV^2 \le q^2 \le 4~GeV^2$. Here we note that the zero--position and
magnitude of ${\cal A}_{\theta_\Lambda^S}^{FB}$ for the 
$\Lambda_b \rar \Lambda \mu^+ \mu^-$ decay are both insensitive to any of
the scalar type interactions.

In Figs. (8), (9) and (10) we present the dependence of ${\cal
A}_{\theta_\Lambda^S}^{FB}$ on $q^2$ at fixed values of $C_{LL}$, $C_{LR}$
and $C_{RL}$, respectively, for the $\Lambda_b \rar \Lambda \tau^+ \tau^-$
decay. Here we would like to note that the dependence of ${\cal
A}_{\theta_\Lambda^S}^{FB}$ on $q^2$ at the given values of $C_{RR}$
coincides practically with that of that of its dependence on $C_{RL}$.
We observe from these figures that, except the resonance region
($q^2\simeq 14.6~GeV^2$), there are no other 
zero--points of ${\cal A}_{\theta_\Lambda^S}^{FB}$ for the Wilson 
coefficients $C_{LL}$ and $C_{RL}$. New zero--points of ${\cal
A}_{\theta_\Lambda^S}^{FB}$ appear in the presence of $C_{LR}$, and they are
being two zero--points at $C_{LR}=-4$ and one zero--point at $C_{LR}=-2$.       
Therefore, determination of the zero--position of ${\cal
A}_{\theta_\Lambda^S}^{FB}$ at $q^2\simeq 17.6~GeV^2$ and $q^2\simeq
18.8~GeV^2$ is an unambiguous indication of the new physics, and this new
physics is solely due to the presence of the Wilson coefficient $C_{LR}$.

It should be noted that ${\cal A}_{\theta_\Lambda^S}^{FB}$ asymmetry for
the $\Lambda_b \rar \Lambda \tau^+ \tau^-$ decay is very sensitive to the
presence of the new scalar type interactions $C_{LRRL}$ and $C_{LRLR}$ (see
Fig. (11)). Since the dependence of ${\cal A}_{\theta_\Lambda^S}^{FB}
(\Lambda_b \rar \Lambda \tau^+ \tau^-)$ on the above--mentioned scalar
coefficients turned out to be very similar, we present the one for the
$C_{LRRL}$ case. From this figure we observe that there appears a new
zero--position of ${\cal A}_{\theta_\Lambda^S}^{FB}$ which is absent in the
SM, and far from the resonance regions considerable departure from the 
SM is predicted. For this reason study of the zero--position of ${\cal
A}_{\theta_\Lambda^S}^{FB}$ can give comprehensive information about the
existence of the new physics beyond the SM.    

We can get additional information by measuring the magnitude of ${\cal
A}_{\theta_\Lambda^S}^{FB}$ in the regions $14.6~GeV^2 \le q^2\le 16~GeV^2$
and $17.6~GeV^2 \le q^2\le 19.2~GeV^2$, which can be useful in determining
the magnitude of the new Wilson coefficients.

In conclusion, we study the dependence of three P--odd forward--backward
asymmetries on $q^2$ in the presence of the new vector type interactions.
The results we obtain can briefly be summarized as follows:

\begin{itemize}

\item The zero--position of ${\cal A}_\theta^{FB}$ for the $\Lambda_b \rar
\Lambda \mu^+ \mu^-$ decay is sensitive only to the presence of $C_{LL}$ and
$C_{LR}$, and is free of the long distance effects. The location of its
zero--position unambiguously allows us to determine the sign of the new
Wilson coefficients.

\item Determination of the value of ${\cal A}_\theta^{FB}$ for the
$\Lambda_b \rar \Lambda \tau^+ \tau^-$ decay between the resonance regions
can give invaluable information about the new physics, which is more
sensitive to the presence of the vector coefficients $C_{RL}$ and $C_{RR}$,
as well as scalar coefficient $C_{LRRL}$.

\item It is shown that the P--odd asymmetry ${\cal A}_{\theta_\Lambda}^{FB}$ 
for the $\Lambda_b \rar \Lambda \mu^+ \mu^-$ and 
$\Lambda_b \rar \Lambda \tau^+ \tau^-$ decays is more sensitive to the
Wilson coefficients $C_{RR}$ and $C_{RL}$, while it is insensitive to the
effects of $C_{LL}$ and $C_{LR}$. In the case of the  $\Lambda_b \rar 
\Lambda \tau^+ \tau^-$ decay, the same asymmetry exhibits strong dependence 
on the scalar coefficients $C_{RLRL}$ and $C_{RLLR}$ as well.

\item Zero--position of ${\cal A}_{\theta_\Lambda^S}^{FB}$ for the
$\Lambda_b \rar \Lambda \mu^+ \mu^-$ decay is practically independent of the
vector type coefficients $C_{LR}$, $C_{RL}$ and $C_{RR}$, and all
type of scalar interactions; but it shows
dependence only on $C_{LL}$. Its zero--point position is shifted to the
right (left) compared to that of the SM result at negative (positive) values
of $C_{LL}$. regions of $q^2$ are found where the value of ${\cal
A}_{\theta_\Lambda^S}^{FB}$ depart from the SM prediction in the presence of
the new vector interactions with the new Wilson coefficients $C_{LR}$,
$C_{RR}$ and $C_{RL}$.

\item Our analysis predicts that, except the resonance regions, there are
new zero--points of ${\cal A}_{\theta_\Lambda^S}^{FB}$ for the negative values
of $C_{RL}$, and for the scalar coefficients $C_{LRRL}$, $C_{LRLR}$ 
for the $\Lambda_b \rar \Lambda \tau^+ \tau^-$ decay.

\end{itemize}

\newpage
 
\section*{Acknowledgments}

One of the authors (T.M.A) is grateful to Prof. Dr. S. Randjbar--Daemi for
the hospitality extended to him at the Abdus Salam International Center for
theoretical Physics, Trieste, Italy, where parts of this work are carried  
out, and to F. Kr\"{u}ger for his collaboration in the early stages of this
work.

\newpage

\section*{Figure captions}
{\bf Fig. (1)} The dependence of the P--odd forward--backward asymmetry 
${\cal A}_\theta^{FB}$ on $q^2$ at five different fixed
values of the vector type Wilson coefficient $C_{LL}$
for the $\Lambda_b \rar \Lambda \mu^+ \mu^-$ decay.\\ \\
{\bf Fig. (2)} The same as in Fig. (1), but for the 
$\Lambda_b \rar \Lambda \tau^+ \tau^-$ decay. \\ \\
{\bf Fig. (3)} The same as in Fig. (2), but at five different fixed
values of the scalar type Wilson coefficient $C_{LRRL}$. \\ \\
{\bf Fig. (4)} The dependence of the P--odd forward--backward asymmetry 
${\cal A}_{\theta_\Lambda}^{FB}$ on $q^2$ at five different fixed
values of the vector type Wilson coefficient $C_{RL}$
for the $\Lambda_b \rar \Lambda \mu^+ \mu^-$ decay.\\ \\
{\bf Fig. (5)} The same as in Fig. (4), but at five different fixed
values of the vector type Wilson coefficient $C_{RR}$. \\ \\
{\bf Fig. (6)} The same as in Fig. (4), but for the            
$\Lambda_b \rar \Lambda \tau^+ \tau^-$ decay. \\ \\
{\bf Fig. (7)} The same as in Fig. (6), but at five different fixed
values of the scalar type Wilson coefficient $C_{RLRL}$. \\ \\
{\bf Fig. (8)} The dependence of the P--odd forward--backward asymmetry 
${\cal A}_{\theta_\Lambda^S}^{FB}$ on $q^2$ at five different fixed
values of the vector type Wilson coefficient $C_{LL}$
for the $\Lambda_b \rar \Lambda \tau^+ \tau^-$ decay.\\ \\
{\bf Fig. (9)} The same as in Fig. (8), but at five different fixed
values of the vector type Wilson coefficient $C_{LR}$. \\ \\
{\bf Fig. (10)} The same as in Fig. (8), but at five different fixed
values of the vector type Wilson coefficient $C_{RL}$.\\ \\
{\bf Fig. (11)} The same as in Fig. (8), but at five different fixed
values of the sclar type Wilson coefficient $C_{LRRL}$.

\newpage

\begin{figure}
\vskip 1.5 cm
    \includegraphics{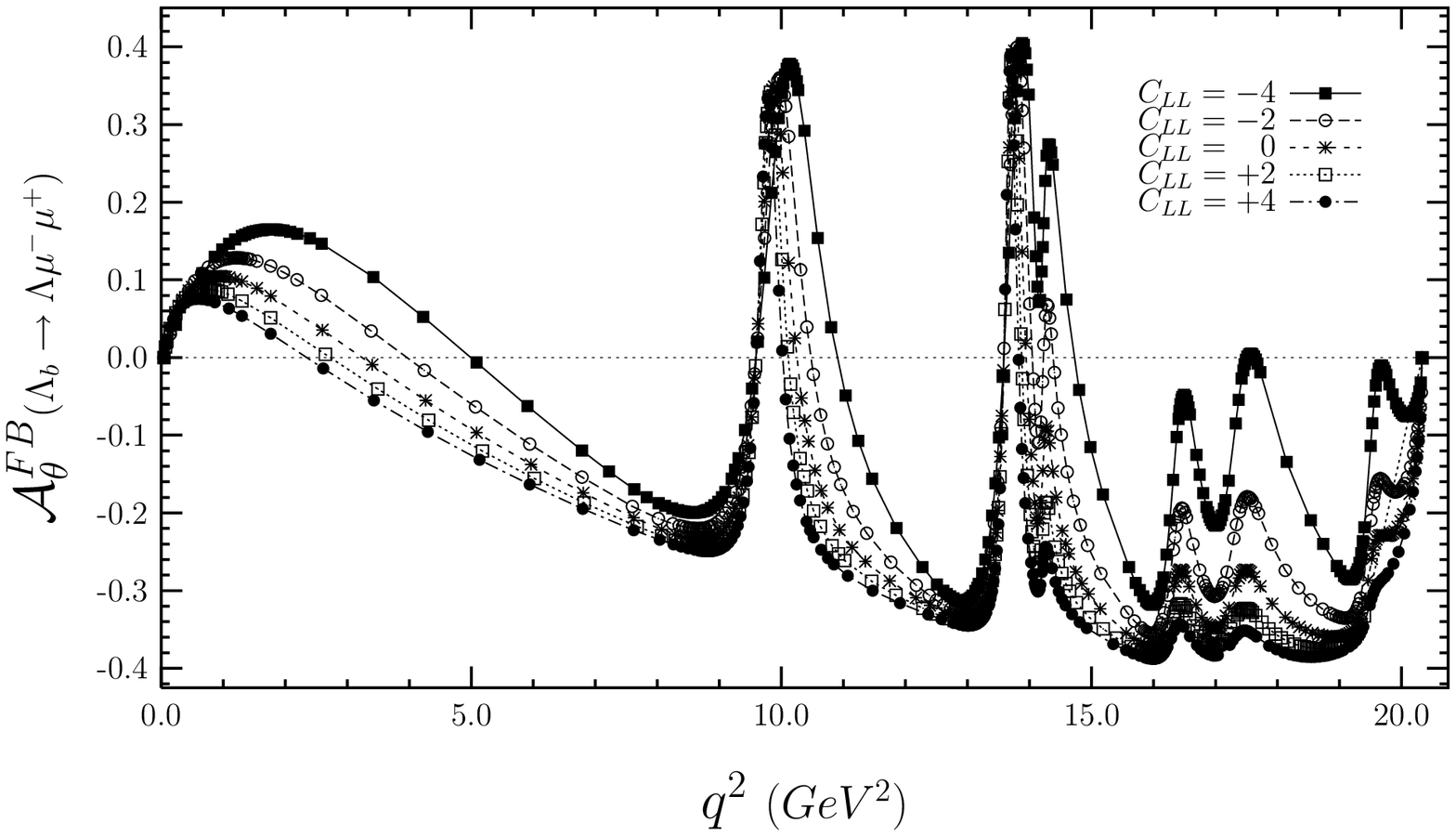}
\vskip 7.8cm
\caption{}
\end{figure}

\begin{figure}
\vskip 2.5 cm
    \includegraphics{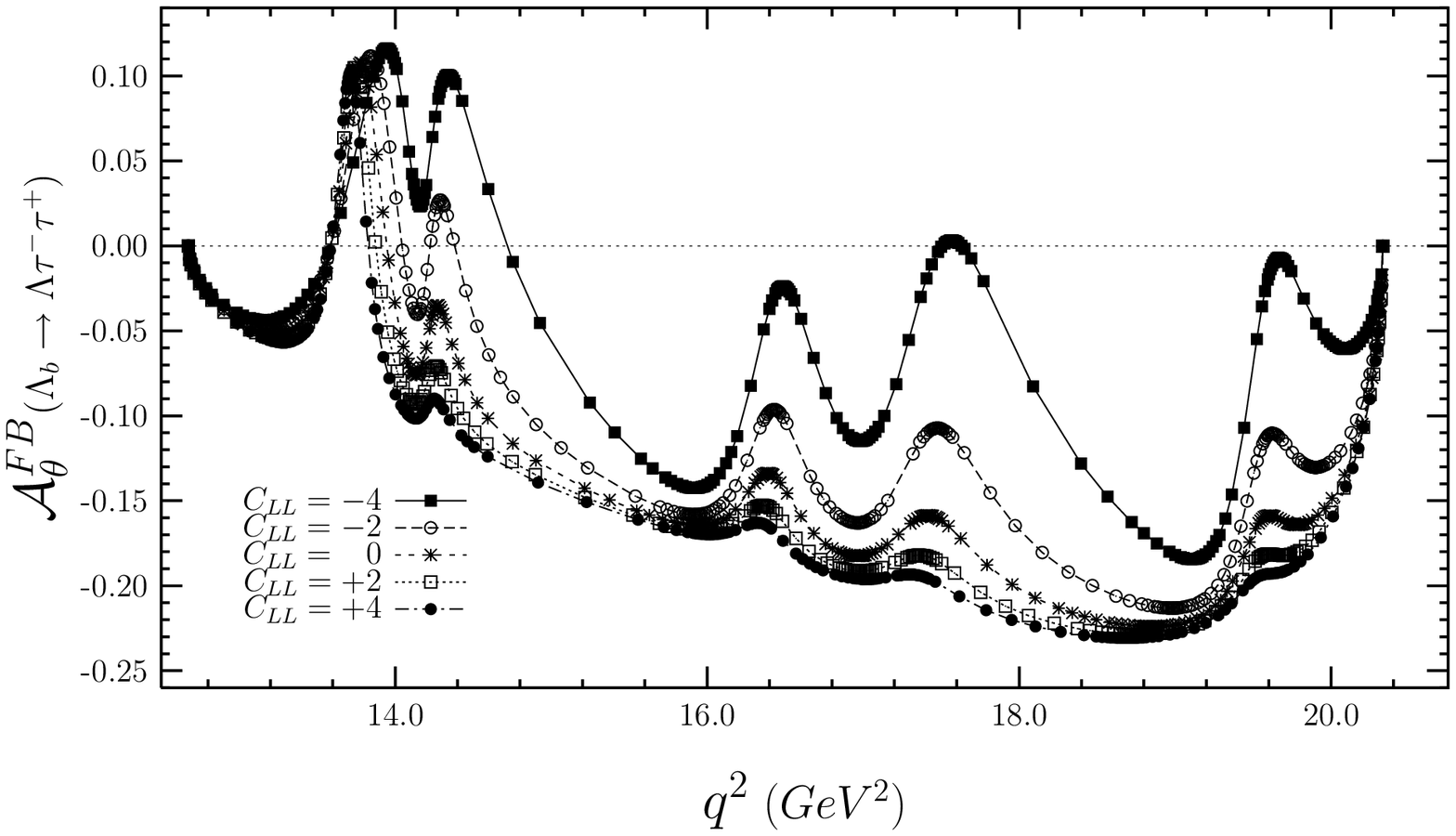}
\vskip 7.8 cm
\caption{}
\end{figure}

\begin{figure}
\vskip 1.5 cm
    \includegraphics{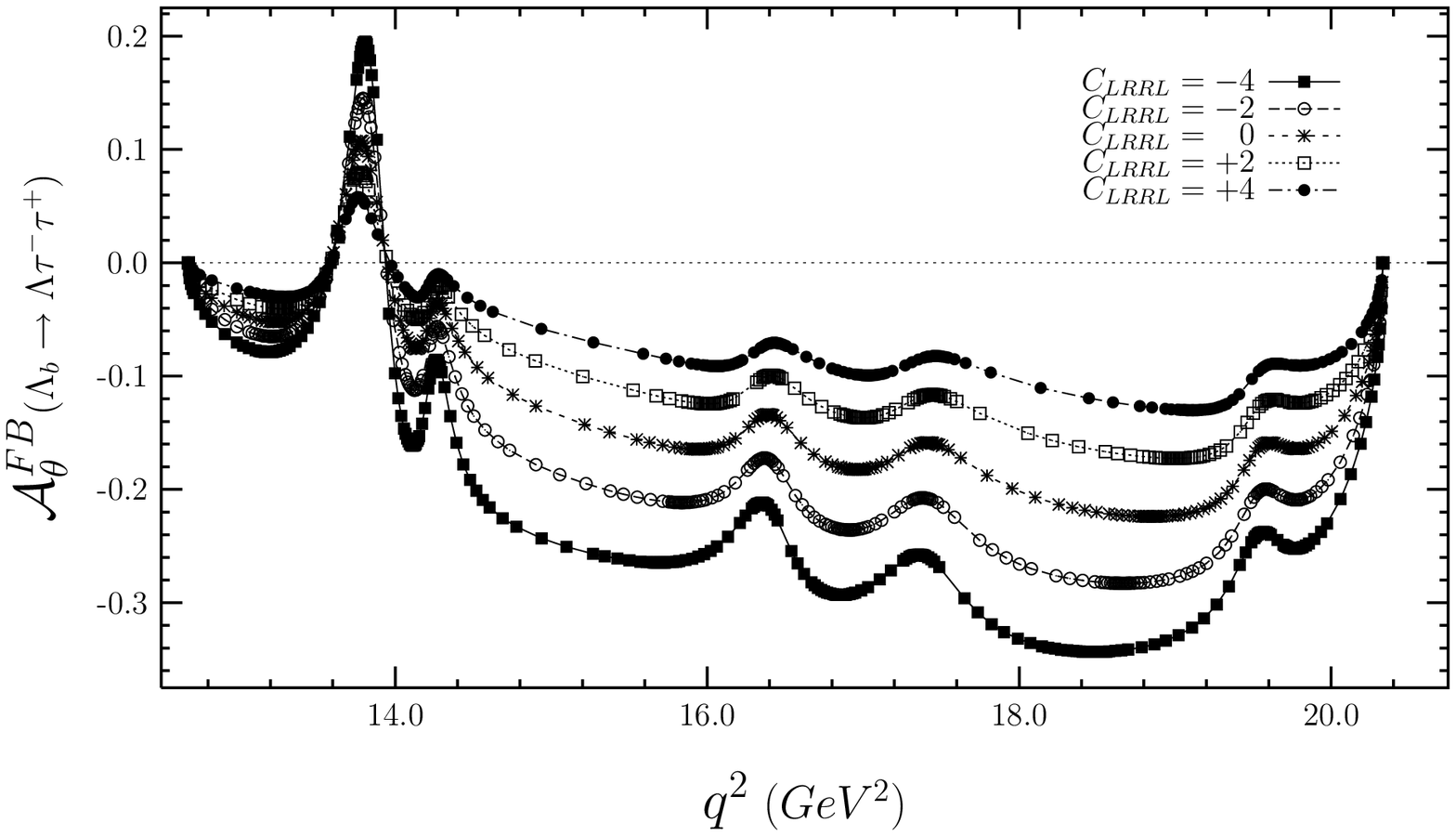}
\vskip 7.8cm
\caption{}
\end{figure}

\begin{figure}
\vskip 2.5 cm
    \includegraphics{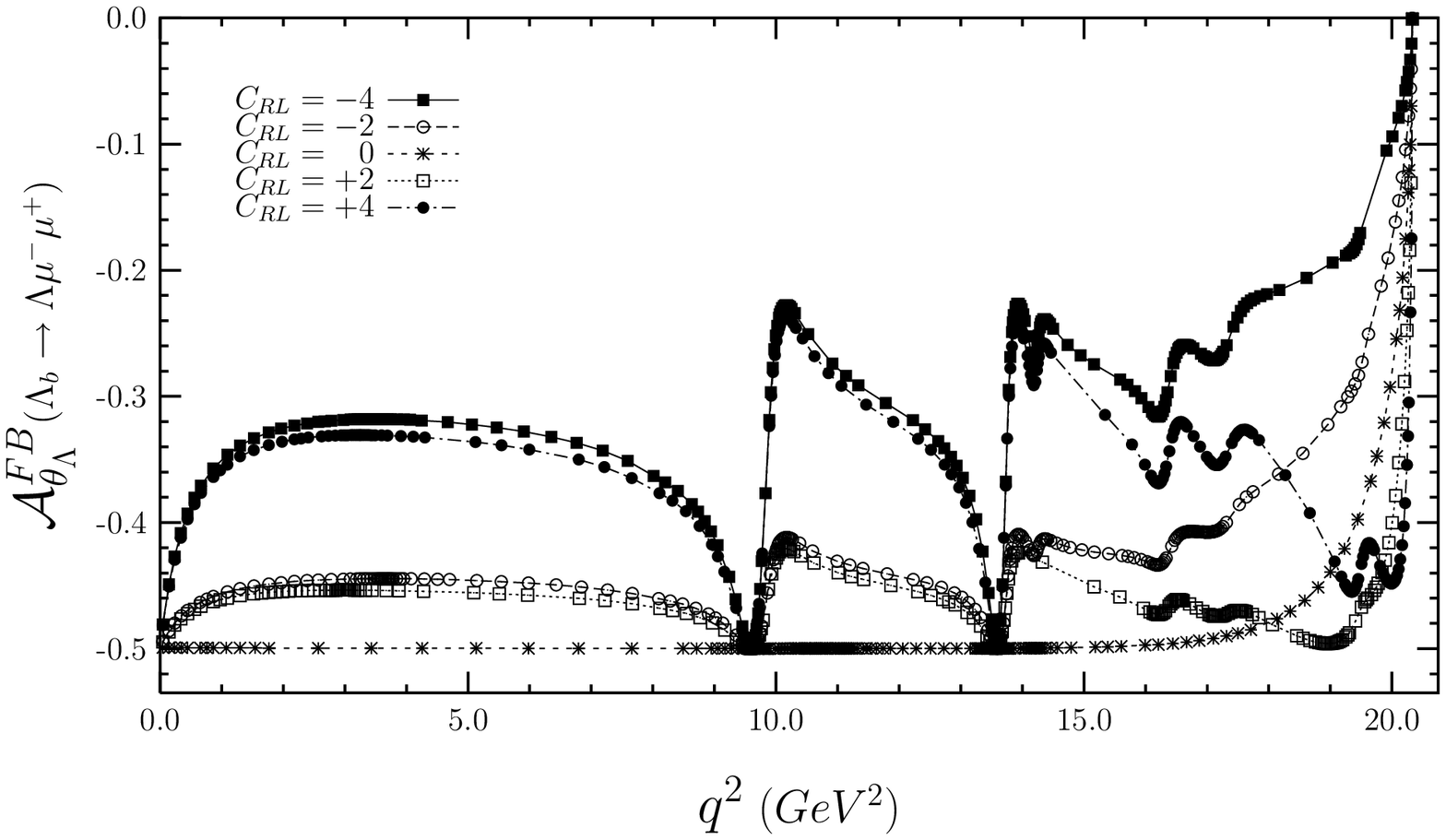}
\vskip 7.8 cm
\caption{}
\end{figure}

\begin{figure}
\vskip 2.5 cm
    \includegraphics{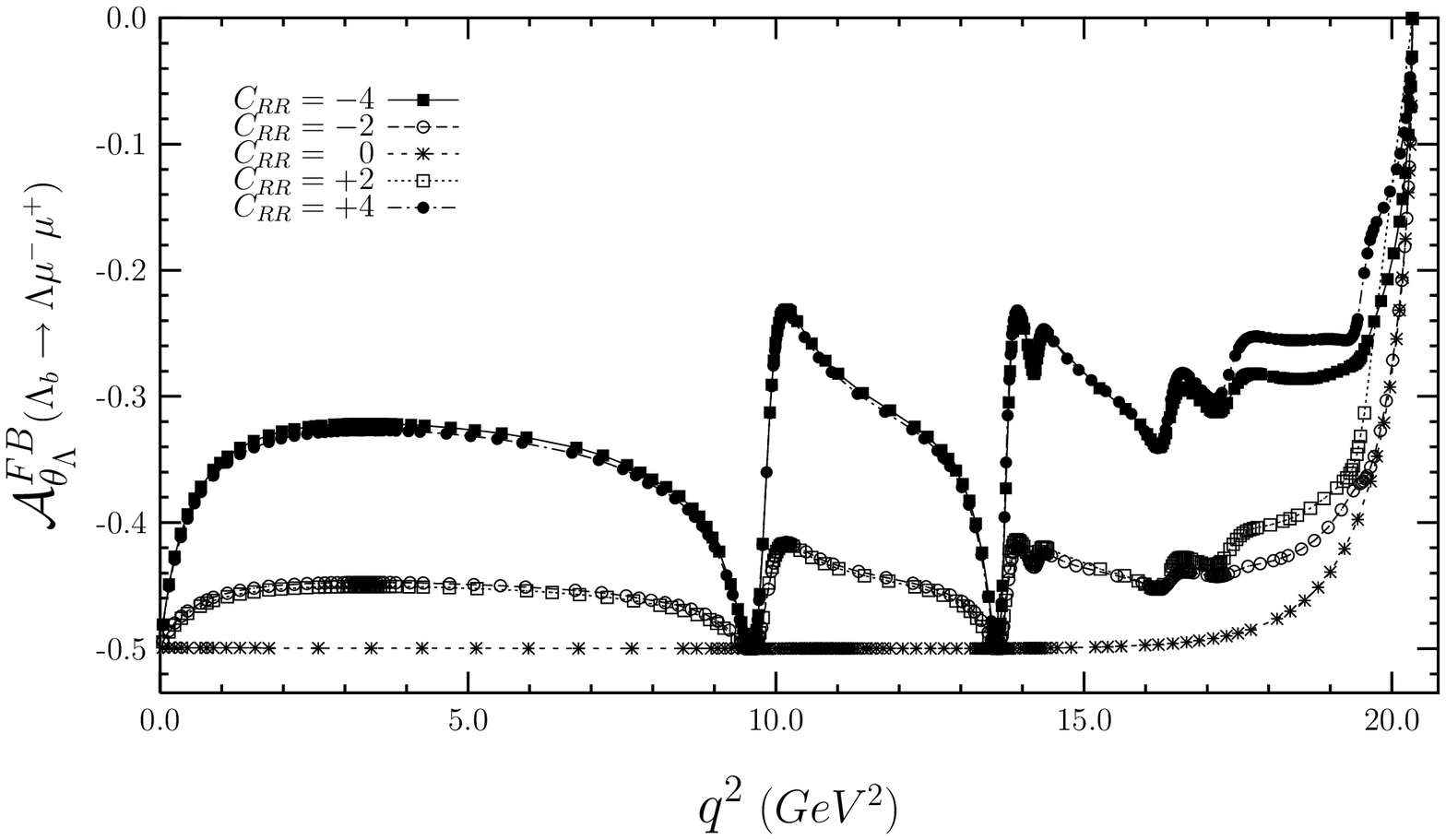}
\vskip 7.8 cm
\caption{}
\end{figure}

\begin{figure}
\vskip 1.5 cm
    \includegraphics{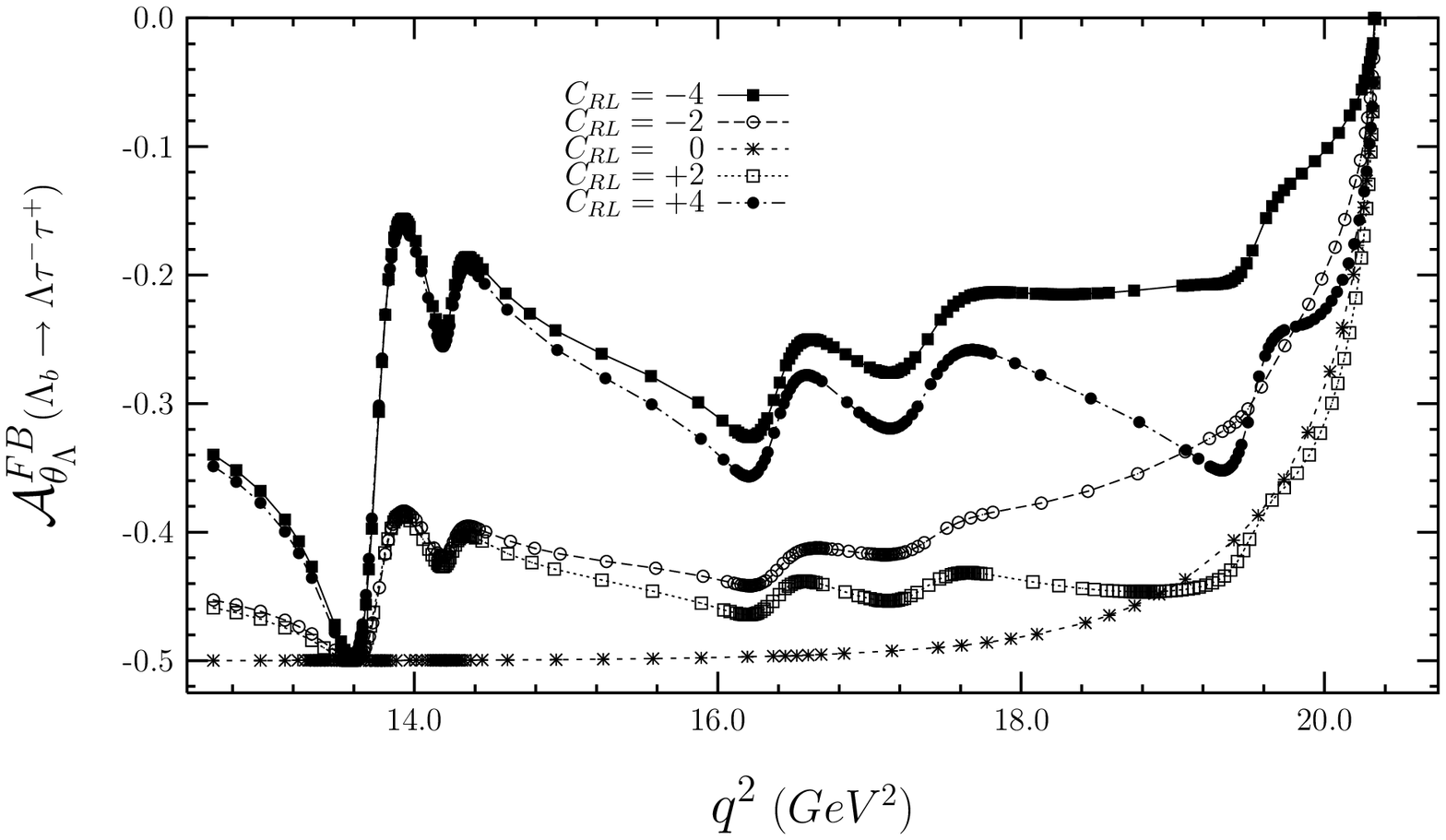}
\vskip 7.8cm
\caption{}
\end{figure}

\begin{figure}
\vskip 2.5 cm
    \includegraphics{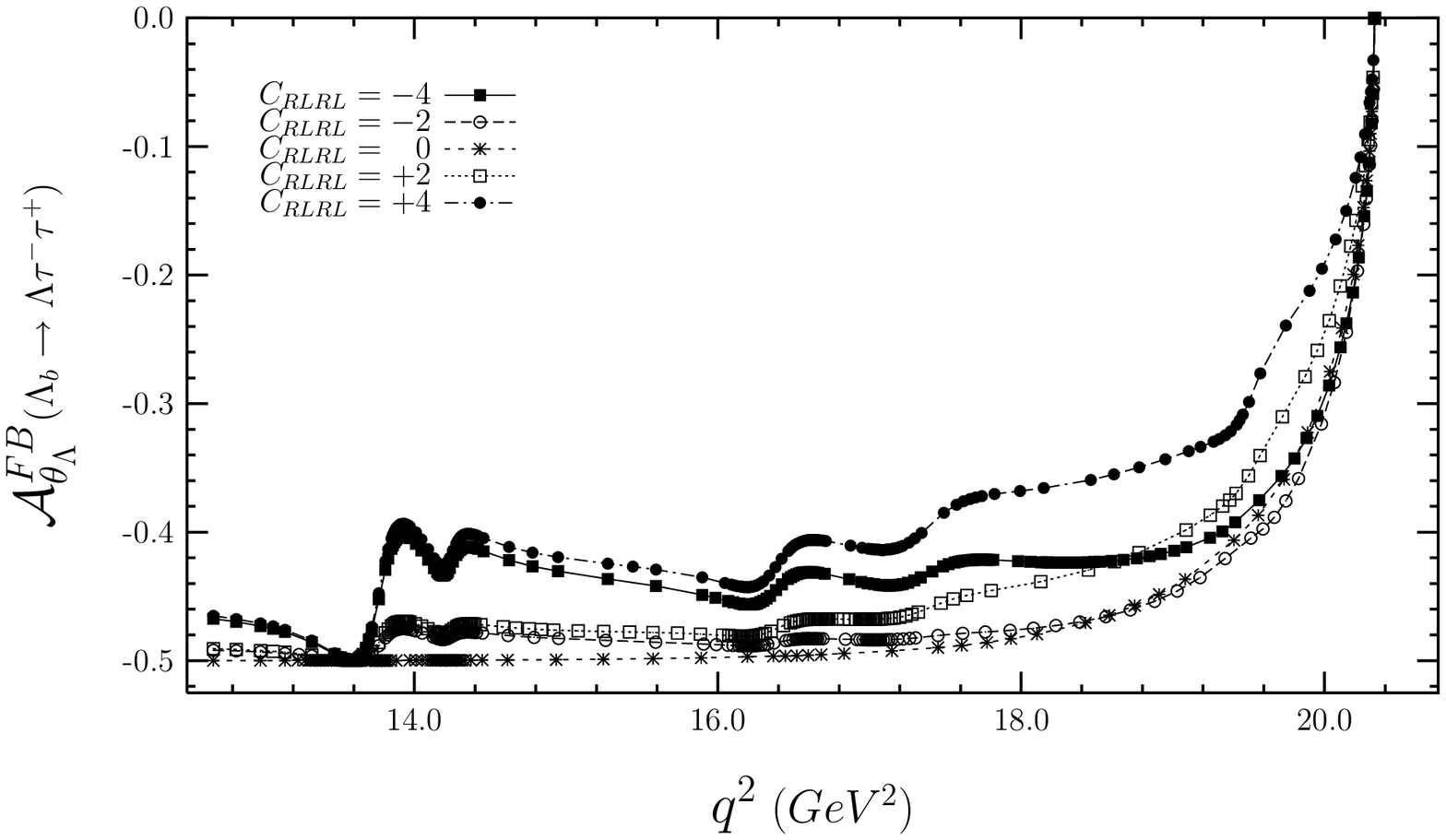}
\vskip 7.8 cm
\caption{}
\end{figure}

\begin{figure}
\vskip 1.5 cm
    \includegraphics{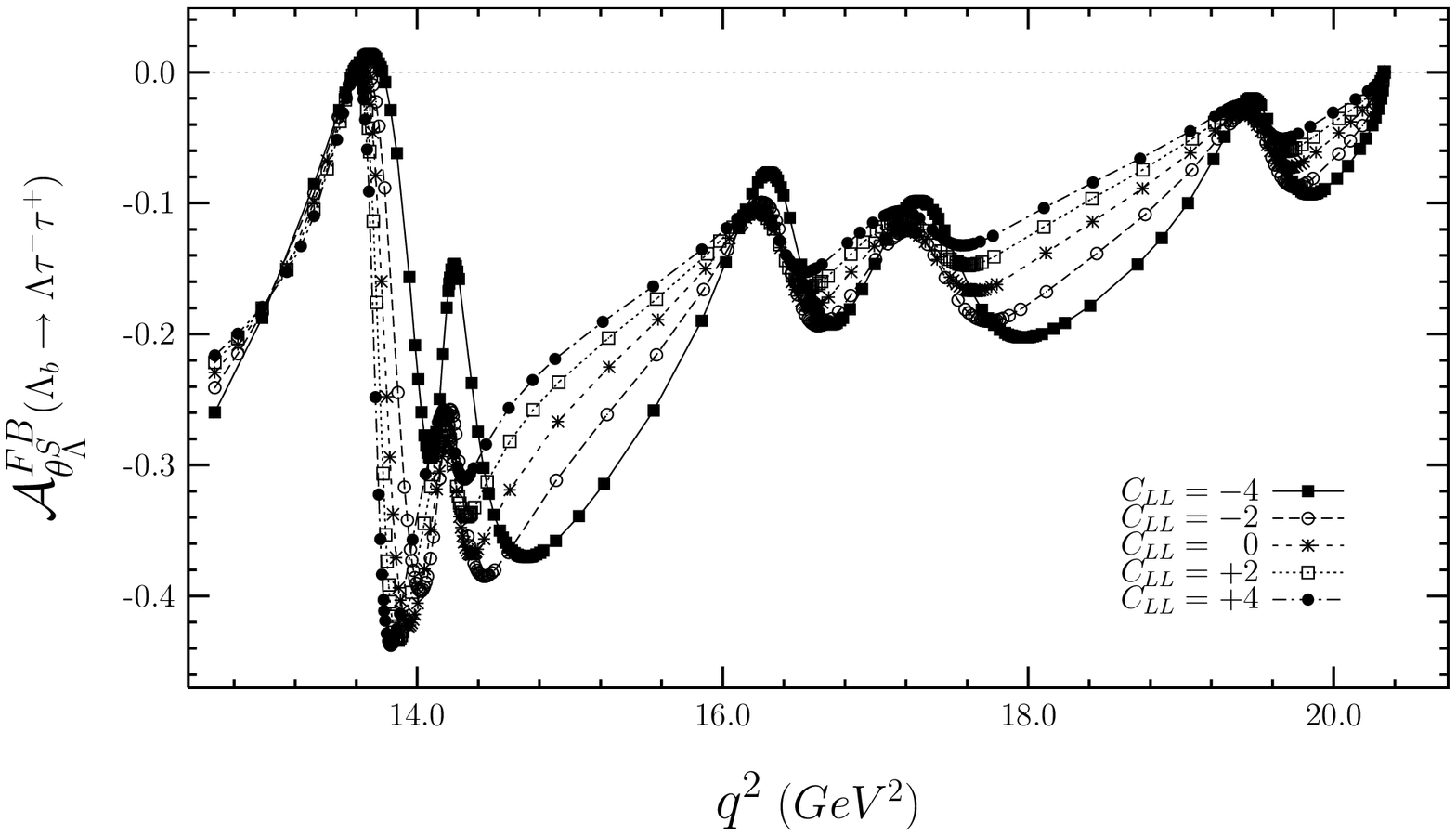}
\vskip 7.8cm
\caption{}
\end{figure}

\begin{figure}
\vskip 2.5 cm
    \includegraphics{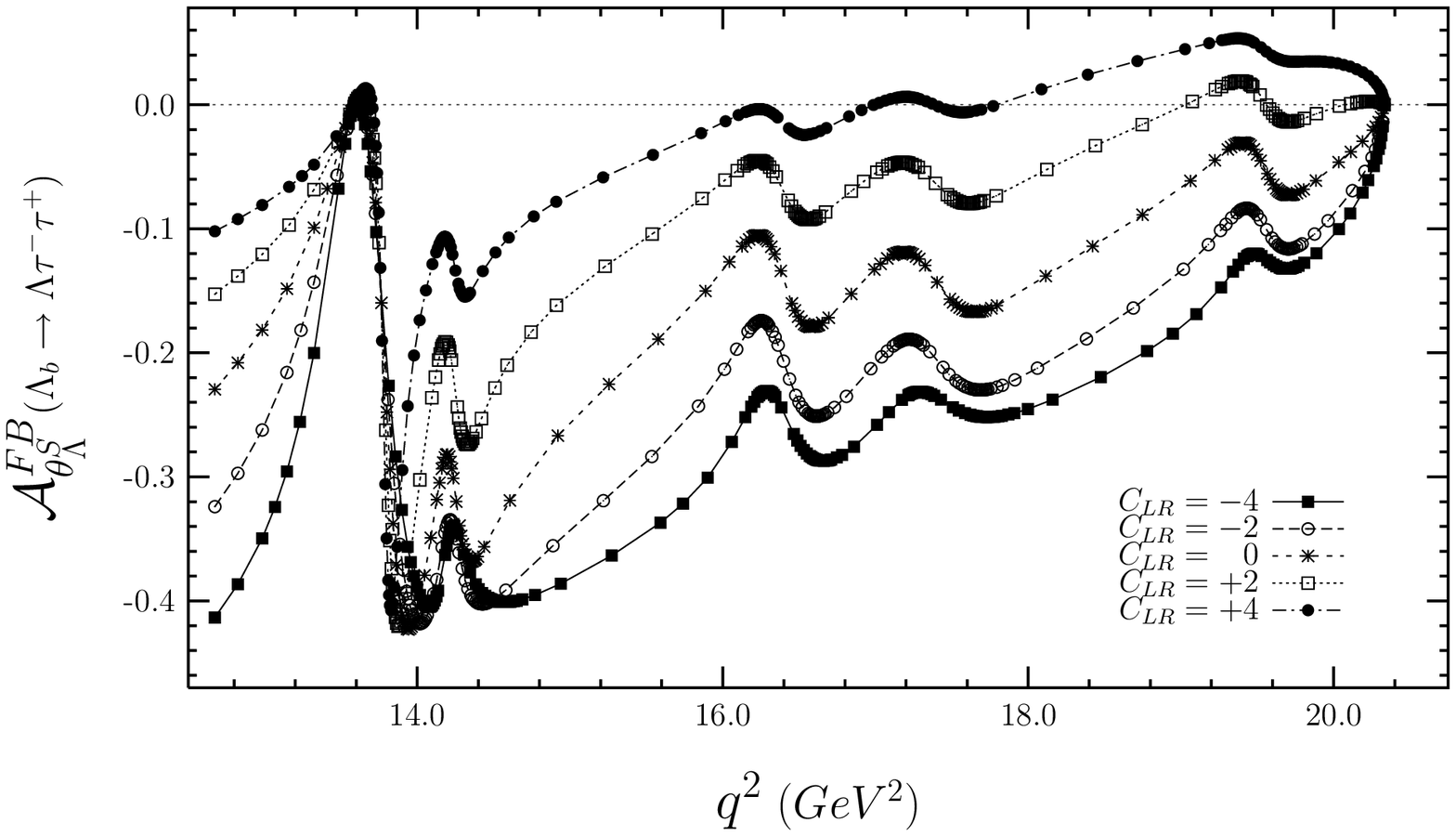}
\vskip 7.8 cm
\caption{}
\end{figure}

\begin{figure}
\vskip 1.5 cm
    \includegraphics{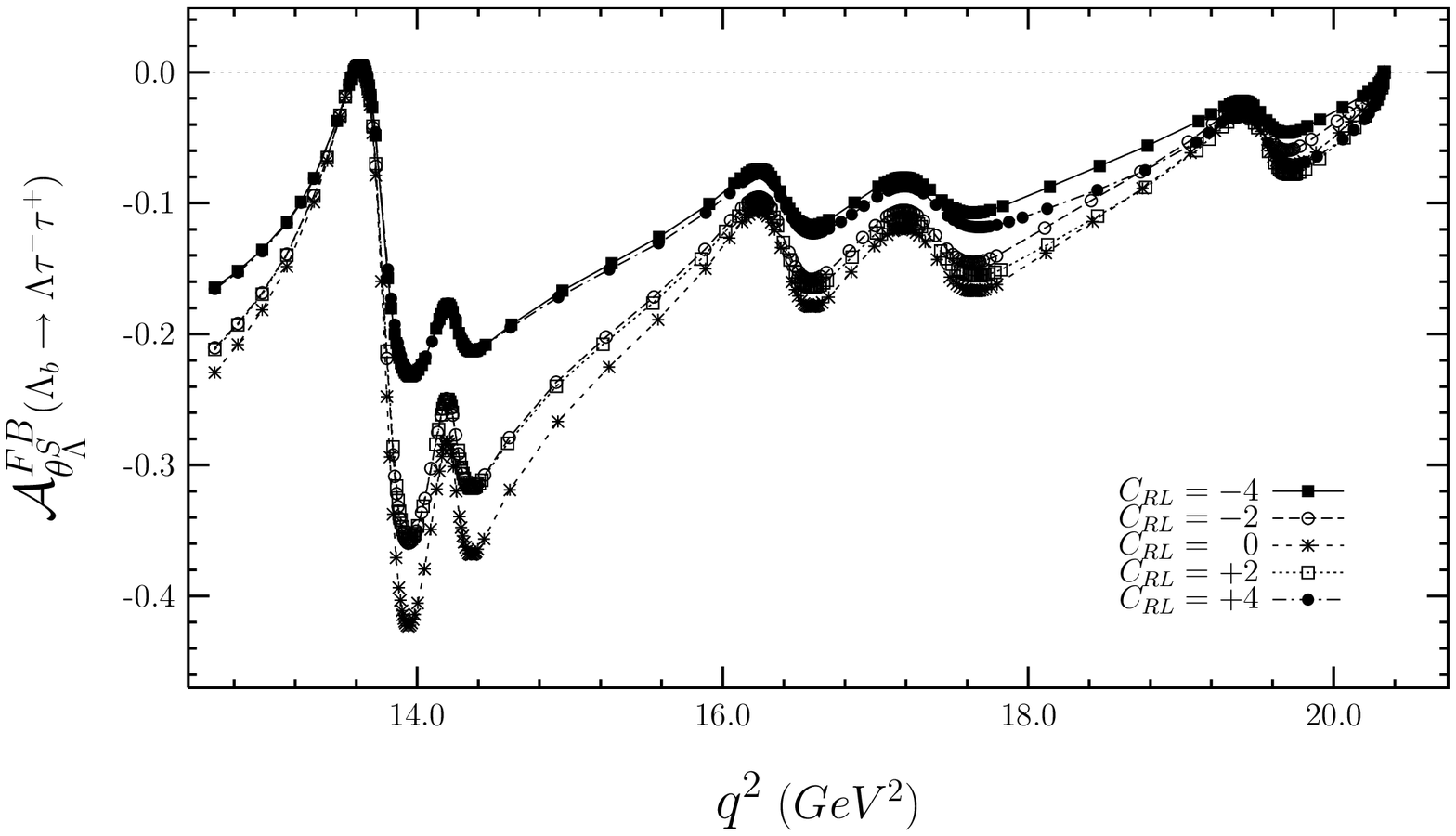}
\vskip 7.8cm
\caption{}
\end{figure}

\begin{figure}
\vskip 2.5 cm
    \includegraphics{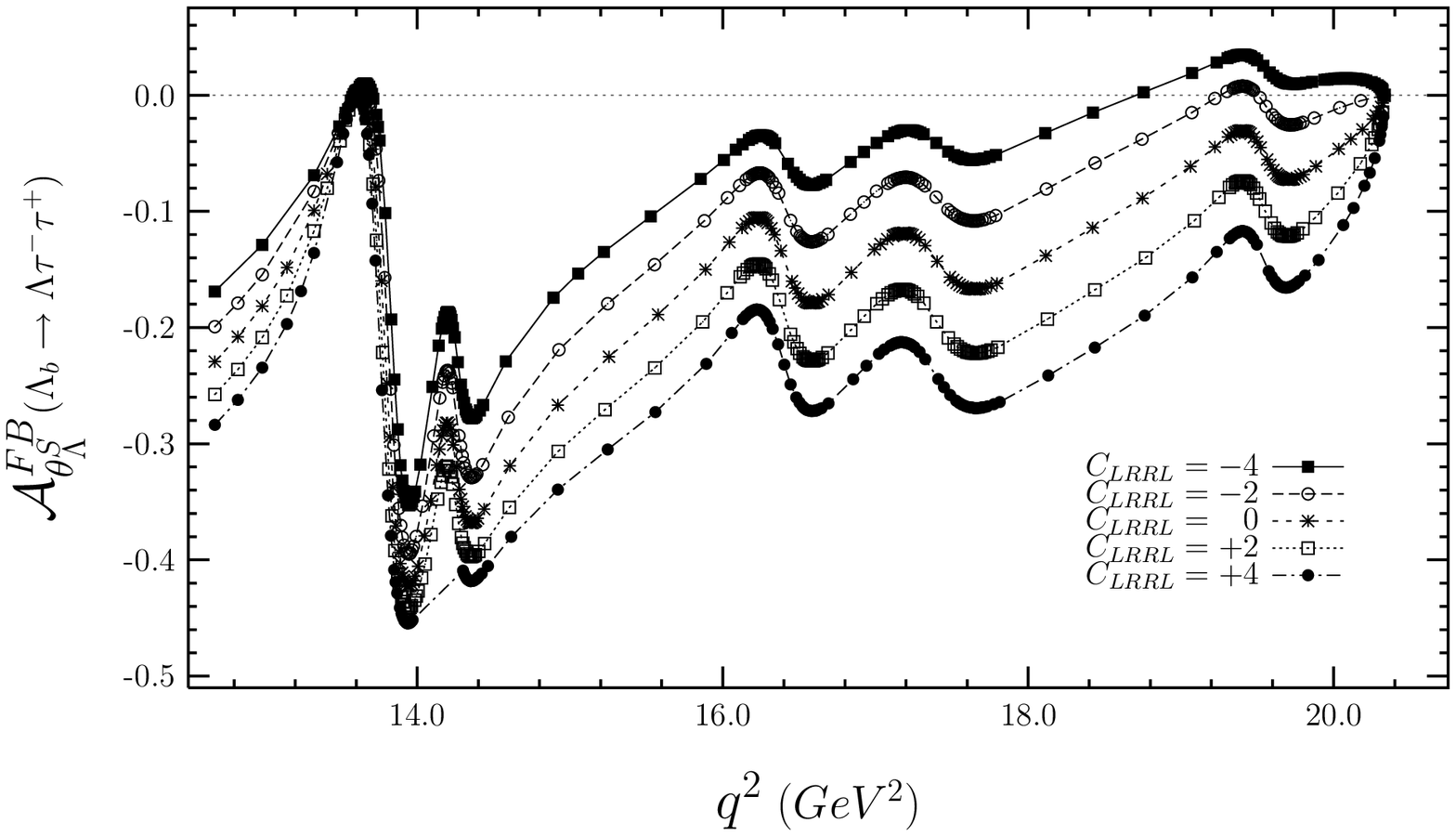}
\vskip 7.8 cm
\caption{}
\end{figure}

\end{document}